# ADAPTIVE SET OBSERVERS DESIGN FOR NONLINEAR CONTINUOUS-TIME SYSTEMS: APPLICATION TO FAULT DETECTION AND DIAGNOSIS

Denis Efimov*, Tarek Raïssi, Ali Zolghadri

*University of Bordeaux, IMS-lab, Automatic control group*
*351 cours de la libération, 33405 Talence, France*
{Denis.Efimov; Tarek.Raıssi; Ali.Zolghadri}@ims-bordeaux.fr

**Abstract** − The paper deals with joint state and parameter estimation for nonlinear continuous-time systems. Based on a guaranteed LPV approximation, the set adaptive observers design problem is solved avoiding the exponential complexity obstruction usually met in the set-membership parameter estimation. Potential application to fault diagnosis is considered. The efficacy of the proposed set adaptive observers is demonstrated on several examples.

## 1. Introduction

The observer design problem for nonlinear systems has been an area of intensive research during the last two decades. There exist a lot of solutions dealing with diverse forms of system models, see for instance [3], [24]. Typically, the observer design problem is solvable if the system model can be transformed to a canonical form, that may be an unacceptable assumption in many applications. Consider a generic nonlinear system

$$\dot{\mathbf{x}} = \mathbf{f}(t, \mathbf{x}, \mathbf{u}, \mathbf{d}) , \quad \mathbf{y} = \mathbf{h}(\mathbf{x}) + \mathbf{v} \tag{1}$$

where $\mathbf{x} \in R^n$, $\mathbf{u} \in R^m$, $\mathbf{d} \in R^l$, $\mathbf{y} \in R^p$, $\mathbf{v} \in R^p$ are respectively the state, the control, the disturbances, the output and the measurement noise; $t \in R$, the functions $\mathbf{f}$, $\mathbf{h}$ are continuous with respect to all arguments and differentiable with respect to $\mathbf{x}$ and $\mathbf{u}$.

In the literature, several observers are built based on an approximation (or a transformation) of the nonlinear model (1) to a Linear Parametric-Varying (LPV) one [6], [19]. LPV models are described by:

$$\dot{\mathbf{x}} = \mathbf{A}(\rho(t))\mathbf{x} + \mathbf{B}(\rho(t))\mathbf{u} , \quad \mathbf{y} = \mathbf{C}(\rho(t))\mathbf{x} + \mathbf{v} , \tag{2}$$

where the scheduling parameter vector $\rho \in \mathcal{P}$ is *a priori* unknown, but with known bounds, and $\mathcal{P}$ is a set of functions that remain in a compact real subspace. Let us stress that the system (2) is an equivalent representation of (1), in the sense that trajectories of (1) remain in the trajectories of (2). Among available methodologies for LPV model constructions one can mention the Jacobian linearization, the state transformation and the state substitution approaches [20], [28], [31]. The idea is to replace nonlinear complexity of the model (1) by enlarged parametric variation in the linear model (2). Such LPV transformation simplifies the design of an observer for the system (1). As it will be shown in this paper, sometimes the complete LPV linearization is not necessary and a partial one may be more suitable. For example, for the observer design purposes some nonlinearities depending only on the output $\mathbf{y}$ can be preserved in order to decrease the uncertainties of the model (2) collected in the vector $\rho$. The observer design methodology proposed in this paper based on a guaranteed LPV transformation recently developed in [26]. By "guaranteed", it is understood that the nonlinear trajectory is sure to remain in the set of trajectories of the resulting LPV model. It is based on an interval linearization around the operational state domain instead of a linearization throughout the equilibrium points. The proposed LPV approximation is performed by means of interval analysis [13], [21].

In the following, an adaptive set observer is developed based on (2) in a set-membership context. There exist three main approaches to perform interval state estimation for systems described by (2): the prediction/correction mechanism

---

as in the Kalman filter [14], [25]; the approach based on comparison theorem [18], [23]; and the closed loop interval observers with cooperative observation error dynamics [2], [12], [22]. The latter has been extended in [26] for nonlinear systems using LPV approximations with known minorant and majorant matrices for (2). Unfortunately, these state estimators are efficient only when the parameter uncertainties are not large.

To the best of our knowledge, joint state and parameter estimation has not been fully studied for systems described by (1) in a bounded error context. An attempt was made in [25] to take into account the uncertain parameters in set-membership framework, where the parameter estimation problem is formulated as a set inversion and solved by the SIVIA algorithm (Set Inversion Via Interval Analysis) [15]. An inclusion test involving a validated integration of a set of ordinary differential equations (ODEs) should be evaluated over a time horizon. Such a procedure is computationally time-consuming since the complexity of SIVIA is *exponential* with respect to the parameter vector dimension. In [16] the validated integration of ODEs is associated with consistency techniques in order to reduce the computing time. Nevertheless, the algorithm in [16] is efficient only for very moderate levels of noise and the complexity remains *exponential*. In the following, the methodology proposed in [26] is extended to deal with joint state and parameter estimation even for higher dimensional systems and with large parametric uncertainty. The idea is to develop set-membership *adaptive* observers based on the works reported in [9], [11], [33], [35].

In this paper a procedure for adaptive set observer design is proposed for a subclass of the LPV representation (2). The main feature of this step is that cooperativity property of the state observers (which can be assigned by the proper choice of the observer gain [26]) is not inherited by the adaptive counterpart. Resolution of this issue requires especial consideration and additional conditions checking. The main advantage is that no bisection is needed in the parameter estimation procedure and the complexity of the algorithm is not exponential. Secondly, a consistency check residual for the nonlinear continuous-time system (1) is computed based on its LPV approximation and the proposed adaptive set observer. Potential application to model based fault diagnosis is then investigated. It is shown that the independently computed estimates of the unknown parameters improve robustness of fault detection, while decreasing the false alarm level.

The paper is organized as follows. In the Section 2 the formal problem statement is presented. Some preliminaries are given in Section 3. The adaptive observer equations and the applicability conditions for the adaptive set observer are derived in Section 4. Two different sets of conditions are analyzed leading to cooperative or competitive adaptive observer loops. The combined set state observer is analyzed in Section 5. Application of the proposed technique to fault detection is considered in Section 6. Through the paper numerical examples are provided to illustrate the results.

## 2. Problem statement

Let us assume that the system (1) can be transformed to the following form:

$$\dot{\mathbf{x}} = \mathbf{A}(\rho(t))\mathbf{x} + \mathbf{B}(\rho(t))\mathbf{u} + \phi(\mathbf{y}) + \mathbf{G}(\mathbf{y})\theta, \qquad (3)$$

$$\mathbf{y} = \mathbf{C}\mathbf{x}, \quad \mathbf{y}_v = \mathbf{y} + \mathbf{v},$$

where $\mathbf{x} \in X \subset R^n$, $\mathbf{u} \in U \subset R^m$, $\mathbf{y} \in Y \subset R^p$ are the state, the input and the output vectors; $\theta \in \Theta \subset R^q$ is the vector of uncertain parameters; $\mathbf{v} \in V \subset R^p$ is the measurement noise; $\mathbf{y}_v$ is the vector of noisy measurements of the system (3), $\rho \in \Upsilon \subset R^r$ is some scheduling parameter vector. The compact sets $X$, $U$, $Y$, $V$, $\Theta$ and $\Upsilon$ are given *a priori*, it assumed that there exist some constant vectors $\mathbf{x}_m, \mathbf{x}_M \in R^n$ such that $\mathbf{x}_m \leq \mathbf{x} \leq \mathbf{x}_M$ for all $\mathbf{x} \in X$. The vector function $\phi$ and columns of the matrix function $\mathbf{G}$ are locally Lipschitz continuous, $\mathbf{C}$ is some constant matrix of appropriate dimension. The majorant matrices $\mathbf{A}_m$, $\mathbf{A}_M$, $\mathbf{B}_m$, $\mathbf{B}_M$ are given such that

$$\mathbf{A}_m \prec \mathbf{A}(\rho) \prec \mathbf{A}_M, \quad \mathbf{B}_m \prec \mathbf{B}(\rho) \prec \mathbf{B}_M$$

for all $\rho \in \Upsilon$ (the inequality $\mathbf{A} \prec \mathbf{B}$ for matrices $\mathbf{A}$, $\mathbf{B}$ with dimension $n \times m$ is understood elementwise $A_{i,j} \leq B_{i,j}$, $i = \overline{1, n}$, $j = \overline{1, m}$). Note, that since $\mathbf{y} \in Y$ and $\mathbf{v} \in V$ there exist constants $k_\phi > 0$, $k_G > 0$ such that $|\phi(\mathbf{y}) - \phi(\mathbf{y}_v)| \leq k_\phi |\mathbf{v}|$ and $|\mathbf{G}(\mathbf{y}) - \mathbf{G}(\mathbf{y}_v)| \leq k_G |\mathbf{v}|$.

**R e m a r k  1**. The output dependency of the function $\mathbf{G}$ as well as the linearity of the output map are the main restrictions on the system (1) and on its LPV transformation. In addition, it is assumed that in the system (3), the LPV transformation is not applied to some nonlinear terms dependent only on the output $\mathbf{y}$ and, the functions $\phi$ and $\mathbf{G}$ are preserved in their original form. In fact, to increase accuracy of the system (1) LPV approximation, one should explicitly handle with care the output dependency in all nonlinearities, thus the most accurate presentation of (3) could be

$$\dot{\mathbf{x}} = \mathbf{A}(\rho(t), \mathbf{y})\mathbf{x} + \mathbf{B}(\rho(t), \mathbf{y})\mathbf{u} + \phi(\mathbf{y}) + \mathbf{G}(\mathbf{y})\boldsymbol{\theta}.$$

In an example below we will consider this issue with more details, however for brevity of presentation, all theoretical results will be formulated only for the system (3) (an extension on the former case is trivial). □

In the following the aim is to design an adaptive observer that, in the noise-free case, provides interval observation of unmeasured components of the state vector $\mathbf{x}$ in (1) and estimates the set of admissible values for the vector $\boldsymbol{\theta}$. For any $\mathbf{v}(t) \in V$, $t \geq 0$ the observer solutions should be bounded.

Finally, parametric fault detection is a potential application of the proposed techniques that is investigated in the last part of the paper. In this case, the vector $\boldsymbol{\theta}$ could be composed of two parts: the first one represents the physical parameters which are not exactly known and the second part contains some "fictive" parameters used to model the effect of faults. The latter parameters (or some of them) become significantly different from their nominal range when a fault occurs. In order to decide whether the detected discrepancy is significant, a decision test, based on a convenient distance, should be used to confirm the presence of a fault. Without loss of generality, the fictive parameters are assumed to have zero value in the nominal fault free case. For a complete fault diagnosis and health monitoring process, this means that some *a priori* knowledge about the faults and their effect is available to build adequately the parameter vector $\boldsymbol{\theta}$ for a given application.

## 3. Preliminaries

*A. Monotone systems*

The system

$$\dot{\mathbf{x}} = \mathbf{f}(t, \mathbf{x}), \quad \mathbf{x} \in X, \quad t \geq 0 \tag{4}$$

with the solution $\mathbf{x}(t, \mathbf{x}_0)$ for the initial condition $\mathbf{x}(0) = \mathbf{x}_0$ is called *monotone*, if $\mathbf{x}_0 \leq \boldsymbol{\xi}_0 \Rightarrow \mathbf{x}(t, \mathbf{x}_0) \leq \mathbf{x}(t, \boldsymbol{\xi}_0)$ for all $t \geq 0$ [29] (for the vectors $\mathbf{x}_0$, $\boldsymbol{\xi}_0$ the inequality $\mathbf{x}_0 \leq \boldsymbol{\xi}_0$ is understood elementwise). The system (4) is called *cooperative* if $\partial f_i(t, \mathbf{x}) / \partial x_j \geq 0$ for all $1 \leq i \neq j \leq n$, $t \in R$ and $\mathbf{x} \in X$ [29]. Cooperative systems form a subclass of monotone ones. A matrix $\mathbf{A}$ with dimension $n \times n$ is called cooperative if $A_{i,j} \geq 0$ for all $1 \leq i \neq j \leq n$. Note that for the cooperative stable system (the matrix $\mathbf{A}$ is cooperative and Hurwitz)

$$\dot{\mathbf{s}}(t) = \mathbf{A}\mathbf{s}(t) + \mathbf{r}(t), \quad \mathbf{s} \in R^n, \quad \mathbf{r} \in R^n, \quad t \geq 0$$

the properties $\mathbf{s}(0) \geq 0$, $\mathbf{r}(t) \geq 0$ for all $t \geq 0$ imply $\mathbf{s}(t) \geq 0$ for $t \geq 0$ and, conversely, $\mathbf{s}(0) \leq 0$, $\mathbf{r}(t) \leq 0$ for all $t \geq 0$ ensures $\mathbf{s}(t) \leq 0$ for $t \geq 0$. The system (4) is called *competitive* if $\partial f_i(t, \mathbf{x}) / \partial x_j \leq 0$ for all $1 \leq i \neq j \leq n$, $t \in R$ and $\mathbf{x} \in X$, the competitive systems behave like cooperative in backward time [29].

*B. Persistency of excitation*

The Lebesgue measurable and square integrable matrix function $\mathbf{R}: R \to R^{l_1 \times l_2}$ with dimension $l_1 \times l_2$ admits $(\ell, \vartheta)$–persistency of excitation (PE) condition, if there exist strictly positive constants $\ell$ and $\vartheta$ such that

$$\int_t^{t+\ell} \mathbf{R}(s) \mathbf{R}(s)^T ds \geq \vartheta \mathbf{I}_{l_1}$$

for any $t \in R$, where $\mathbf{I}_{l_1}$ denotes identity matrix of dimension $l_1 \times l_1$ [1], [34].

**Lemma 1** [9]. *Consider the time-varying linear dynamical system*

$$\dot{\mathbf{p}} = -\Gamma \mathbf{R}(t) \mathbf{R}(t)^T \mathbf{p} + \mathbf{b}(t), \ t_0 \in R_+,$$

*where $\mathbf{p} \in R^{l_1}$, $\Gamma$ is a positive definite symmetric matrix of dimension $l_1 \times l_1$ and the functions $\mathbf{R}: R_+ \to R^{l_1 \times l_2}$, $\mathbf{b}: R_+ \to R^{l_1}$ are Lebesgue measurable, $\mathbf{b}$ is essentially bounded, function $\mathbf{R}$ is $(\ell, \vartheta)$–PE for some $\ell > 0$, $\vartheta > 0$. Then, for any initial condition $\mathbf{p}(t_0) \in R^{l_1}$, the solution of the system is defined for all $t \geq t_0$ and verifies ($\gamma > 0$ is the smallest eigenvalue of the matrix $\Gamma$)*

$$|\mathbf{p}(t)| \leq |\mathbf{p}(t_0)| e^{-0.5 \gamma \vartheta \ell^{-1}(t - t_0 - \ell)} + (1 + 2\vartheta^{-1} \gamma^{-1} e^{-0.5 \vartheta \gamma}) \ell \|\mathbf{b}\|. \qquad \square$$

This lemma states that a linear system with a persistently excited time-varying matrix gain and a bounded additive disturbance has bounded solutions.

## 4. Interval parameters estimation

To proceed, we would like to introduce the following assumptions dealing with stabilizability by output feedback of the system (3) linear part.

**Assumption 1**. *There exist matrices $\mathbf{L}$, $\mathbf{Q} = \mathbf{Q}^T > 0$ and $\mathbf{P} = \mathbf{P}^T > 0$ such that*

$$[\mathbf{A}(\rho) - \mathbf{LC}]^T \mathbf{P} + \mathbf{P}[\mathbf{A}(\rho) - \mathbf{LC}] = -\mathbf{Q}$$

*for all $\rho \in \Upsilon$.* $\qquad \square$

For the system

$$\dot{\mathbf{s}} = [\mathbf{A}(\rho(t)) - \mathbf{LC}]\mathbf{s} + \mathbf{r}, \ \mathbf{s} \in R^n, \ \mathbf{r} \in R^n, \ \rho(t) \in \Upsilon \ \text{for} \ t \geq 0, \qquad (5)$$

this assumption ensures uniform asymptotic stability property for $\mathbf{r} = 0$ and boundedness of the system solutions for any bounded input $\mathbf{r}$ (input-to-state stability property holds [30]). The system (5) is the linear part of (3) closed by output feedback with a gain $\mathbf{L}$. This assumption is required for classical adaptive observer design for the system (3). It will be shown later that this assumption is not actually required for the proposed approach. It will be relaxed leading to the following assumption, that ensures existence of an adaptive set observer for (3).

**Assumption 2**. *There exist matrices $\mathbf{L}_m$, $\mathbf{L}_M$ such that the matrices $\mathbf{A}_m - \mathbf{L}_m \mathbf{C}$ and $\mathbf{A}_M - \mathbf{L}_M \mathbf{C}$ are Hurwitz and cooperative, and for all $\mathbf{y} \in Y$, $\mathbf{v} \in V$ we have $0 \prec \mathbf{G}(\mathbf{y} + \mathbf{v})$.* $\qquad \square$

In addition, since $\theta \in \Theta$, there exist two vectors $\theta_m \in R^q$ and $\theta_M \in R^q$ such that $\theta_m \leq \theta \leq \theta_M$ for all $\theta \in \Theta$. Based on these assumptions, the equations of adaptive observer are introduced below in two steps.

*A. The ideal case*

Firstly, assume that the signal $\rho(t) \in \Upsilon$ is available for measurements and assumption 1 holds. Then, an adaptive observer [33], [35] for the system (3) could be built as:

$$\dot{\zeta} = \mathbf{A}(\rho(t))\zeta + \mathbf{B}(\rho(t))\mathbf{u} + \phi(\mathbf{y}_v) + \mathbf{L}(\mathbf{y}_v - \mathbf{C}\zeta);  \qquad (6)$$

$$\dot{\mathbf{\Omega}} = [\mathbf{A}(\rho(t)) - \mathbf{LC}]\mathbf{\Omega} - \mathbf{G}(\mathbf{y}_v); \qquad (7)$$

$$\dot{\hat{\mathbf{\theta}}} = -\mathbf{\Gamma}_0 \mathbf{\Omega}^T \mathbf{C}^T (\mathbf{y}_v - \mathbf{C}\zeta + \mathbf{C}\mathbf{\Omega}\hat{\mathbf{\theta}}), \quad \mathbf{\Gamma}_0 = \mathbf{\Gamma}_0^T > 0, \qquad (8)$$

where $\zeta \in R^n$ is the vector of "estimates" for $\mathbf{x}$; the matrix $\mathbf{\Omega} \in R^{n \times q}$ is an auxiliary variable, which helps to overcome high relative degree obstruction in the system (3), i.e. to identify the value of $\mathbf{\theta}$ even in the cases when only higher order time derivatives of the output $\mathbf{y}$ depend on $\mathbf{\theta}$; $\hat{\mathbf{\theta}} \in R^q$ is the estimate of $\mathbf{\theta}$. Defining the observation error $\varepsilon = \mathbf{x} - \zeta$, the estimation error $\tilde{\mathbf{\theta}} = \mathbf{\theta} - \hat{\mathbf{\theta}}$ and the auxiliary variable $\delta = \varepsilon + \mathbf{\Omega}\mathbf{\theta}$ we obtain

$$\dot{\varepsilon} = [\mathbf{A}(\rho(t)) - \mathbf{LC}]\varepsilon + \mathbf{G}(\mathbf{y}_v)\mathbf{\theta} + \mathbf{d}_v, \qquad (9)$$

$$\mathbf{d}_v = \phi(\mathbf{y}) - \phi(\mathbf{y}_v) + [\mathbf{G}(\mathbf{y}) - \mathbf{G}(\mathbf{y}_v)]\mathbf{\theta} - \mathbf{L}\mathbf{v},$$

$$\dot{\delta} = [\mathbf{A}(\rho(t)) - \mathbf{LC}]\delta + \mathbf{d}_v, \qquad (10)$$

$$\dot{\tilde{\mathbf{\theta}}} = \mathbf{\Gamma}_0 \mathbf{\Omega}^T \mathbf{C}^T (\mathbf{C}\delta + \mathbf{v} - \mathbf{C}\mathbf{\Omega}\tilde{\mathbf{\theta}}). \qquad (11)$$

As in [9], [11], [33], [35], if assumption 1 is satisfied and $\mathbf{y} \in Y$, $\mathbf{v} \in V$, then since the systems (7) and (10) have form similar to (5), all solutions of the system (7) are bounded, i.e. there exists $k_\Omega > 0$ such that $|\mathbf{\Omega}(t)| \leq k_\Omega$ for all $t \geq 0$. Furthermore, we have that $|\mathbf{d}_v| \leq [k_\phi + k_G |\mathbf{\theta}| + |\mathbf{L}|]|\mathbf{v}|$ for $\mathbf{\theta} \in \Theta$, $\mathbf{v} \in V$, then the signal $\mathbf{d}_v$ remains bounded with amplitude proportional to that of $\mathbf{v}$. Therefore, the solutions of (10) are bounded and for the case $\mathbf{v}(t) = 0$, $t \geq 0$ the system is asymptotically stable. In addition, if the signal $\mathbf{\Omega}^T(t)\mathbf{C}^T$ is persistently exciting, then from lemma 1 the estimation error $\tilde{\mathbf{\theta}}(t)$ remains bounded, and for $\mathbf{v}(t) = 0$, $t \geq 0$ the asymptotic relation holds:

$$\lim_{t \to +\infty} \hat{\mathbf{\theta}}(t) = \mathbf{\theta}.$$

Finally, $\varepsilon(t) = \delta(t) - \mathbf{\Omega}(t)\mathbf{\theta}$ for all $t \in R$ and the observation error is bounded since the signals $\delta(t)$ and $\mathbf{\Omega}(t)$ have the same boundedness property. Therefore, the system (6)–(8) is an estimator for $\mathbf{\theta}$ in the noise free case. The presence of noise does not destabilize the observer. Note, that as in [9], [11], [33], [35] a complication of the equation (6) allows one to ensure observation of $\mathbf{x}(t)$ by $\zeta(t)$, however, as it will be shown later such a nice property is not inherited by an adaptive set observer. This is why the simplified equation (6) is considered here.

Moreover, since the system (7) is a stable time-varying filter, the requirement that the signal $\mathbf{C}^T\mathbf{\Omega}^T(t)$ should be PE is related with the same properties of the signal $\mathbf{G}(\mathbf{y}_v(t))$.

*B. The adaptive set observer equations*

Usually the signal $\rho(t) \in \Upsilon$ is not measured and not available on-line, thus the observer (6)–(8) is not realizable. For this case we propose an interval observer based on Assumption 2 instead of Assumption 1 previously:

$$\dot{\zeta}_o = \mathbf{A}_o \zeta_o + \mathbf{B}_o \mathbf{u} + \phi(\mathbf{y}_v) + \mathbf{L}_o(\mathbf{y}_v - \mathbf{C}\zeta_o); \qquad (12)$$

$$\dot{\mathbf{\Omega}}_o = [\mathbf{A}_o - \mathbf{LC}]\mathbf{\Omega}_o - \mathbf{G}(\mathbf{y}_v); \qquad (13)$$

$$\dot{\hat{\mathbf{\theta}}}_o = -\mathbf{\Gamma}_o \mathbf{\Omega}_o^T \mathbf{C}^T (\mathbf{y}_v - \mathbf{C}\zeta_o + \mathbf{C}\mathbf{\Omega}_o \hat{\mathbf{\theta}}_o), \qquad (14)$$

where the index $o \in \{m, M\}$ denotes the upper and lower interval bounds, $\zeta_o \in R^n$, $\mathbf{\Omega}_o \in R^{n \times q}$ and $\hat{\mathbf{\theta}}_o \in R^q$ have the same meaning, the matrix $\mathbf{\Gamma}_o = \mathbf{\Gamma}_o^T > 0$ is a design parameter of the algorithm (14).

In set observer design the monotonicity property of observers equations plays an essential role. As it can be deduced from equations (12)−(14), the monotonicity of the first two subsystems (12), (13) is predefined by assumption 2 conditions. Monotonicity of the system (14), that defines dynamics of parameters estimator, may not be followed by the same property of the systems (12), (13). Actually, it is shown below that under some conditions, the dynamics of the system (14) can be either *cooperative* or *competitive*, impacting the admissible set of $\boldsymbol{\theta}$ construction. In the following subsections each case will be analyzed and the new results are summarized in the theorems 1 and 2.

*C. The competitive case*

The following theorem establishes stability and monotonicity properties of the observers (12)−(14) for $o \in \{m, M\}$.

**T h e o r e m  1**. *Let assumption 2 hold, and $\mathbf{x}(t) \in X$, $\mathbf{u}(t) \in U$, $\mathbf{v}(t) \in V$, $\rho(t) \in \Upsilon$ and $\boldsymbol{\theta} \in \Theta$ for all $t \geq 0$, and assume that the signals $\boldsymbol{\Omega}_o^T(t)\mathbf{C}^T$ are $(\ell_o, \vartheta_o)$–PE for some $\ell_o > 0$, $\vartheta_o > 0$, $o \in \{m, M\}$. Then*:

(i) *for all $t \geq 0$ and $o \in \{m, M\}$ the solutions $\zeta_o(t)$, $\boldsymbol{\Omega}_o(t)$ and $\hat{\boldsymbol{\theta}}_o(t)$ of the system (12)−(14) are bounded provided that $\mathbf{v}(t) \in V$, $t \geq 0$*;

(ii) $0 \prec \mathbf{C}$, $\mathbf{v}(t) \equiv 0$ *for all $t \geq 0$ and there exists a matrix $\overline{\mathbf{\Gamma}}$ such that for all $0 \prec \mathbf{\Gamma}_o \prec \overline{\mathbf{\Gamma}}$, $o \in \{m, M\}$*,

  a. *if $\boldsymbol{\Omega}_o(0) = 0$, $o \in \{m, M\}$, $\boldsymbol{\varepsilon}_m(0) \geq 0$, $\boldsymbol{\varepsilon}_M(0) \leq 0$, $\hat{\boldsymbol{\theta}}_M(0) = \boldsymbol{\theta}_m$, $\hat{\boldsymbol{\theta}}_m(0) = \boldsymbol{\theta}_M$ and there exist*
  $$\mathbf{b}_o = -\lim_{T \to +\infty} T^{-1} \int_0^T \boldsymbol{\Omega}_o^T(t)\mathbf{C}^T \mathbf{C} \boldsymbol{\varepsilon}_o(t) dt, \quad \mathbf{R}_o = \lim_{T \to +\infty} T^{-1} \int_0^T \boldsymbol{\Omega}_o^T(t) \mathbf{C}^T \mathbf{C} \boldsymbol{\Omega}_o(t) dt, \quad o \in \{m, M\} \text{ such that }$$
  $\boldsymbol{\theta}_M < \mathbf{R}_m^{-1} \mathbf{b}_m$, $\mathbf{R}_M^{-1} \mathbf{b}_M < \boldsymbol{\theta}_m$, *then* $\hat{\boldsymbol{\theta}}_M(t) \leq \boldsymbol{\theta} \leq \hat{\boldsymbol{\theta}}_m(t)$, $t \geq 0$.

  b. *if $\boldsymbol{\Omega}_o(0) = 0$, $o \in \{m, M\}$, $\boldsymbol{\varepsilon}_m(0) \leq 0$, $\boldsymbol{\varepsilon}_M(0) \geq 0$, $\hat{\boldsymbol{\theta}}_m(0) = \boldsymbol{\theta}_m$, $\hat{\boldsymbol{\theta}}_M(0) = \boldsymbol{\theta}_M$ and $\boldsymbol{\theta}_M < \mathbf{R}_M^{-1}\mathbf{b}_M$, $\mathbf{R}_m^{-1}\mathbf{b}_m < \boldsymbol{\theta}_m$, then $\hat{\boldsymbol{\theta}}_m(t) \leq \boldsymbol{\theta} \leq \hat{\boldsymbol{\theta}}_M(t)$, $t \geq 0$.*

**P r o o f**. Define $\boldsymbol{\varepsilon}_o = \mathbf{x} - \boldsymbol{\zeta}_o$, $\tilde{\boldsymbol{\theta}}_o = \boldsymbol{\theta} - \hat{\boldsymbol{\theta}}_o$ and $\boldsymbol{\delta}_o = \boldsymbol{\varepsilon}_o + \boldsymbol{\Omega}_o \boldsymbol{\theta}$ for $o \in \{m, M\}$, then we obtain

$$\dot{\boldsymbol{\varepsilon}}_o = [\mathbf{A}_o - \mathbf{L}_o \mathbf{C}]\boldsymbol{\varepsilon}_o + \mathbf{G}(\mathbf{y}_v)\boldsymbol{\theta} + \mathbf{p}_o + \mathbf{d}_v, \quad \mathbf{p}_o = [\mathbf{A}(\rho(t)) - \mathbf{A}_o]\mathbf{x} + [\mathbf{B}(\rho(t)) - \mathbf{B}_o]\mathbf{u}, \qquad (15)$$

$$\dot{\boldsymbol{\delta}}_o = [\mathbf{A}_o - \mathbf{L}_o \mathbf{C}]\boldsymbol{\delta}_o + \mathbf{p}_o + \mathbf{d}_v, \qquad (16)$$

$$\dot{\hat{\boldsymbol{\theta}}}_o = \boldsymbol{\Gamma}_o \boldsymbol{\Omega}_o^T \mathbf{C}^T (\mathbf{C}\boldsymbol{\delta}_o + \mathbf{v} - \mathbf{C}\boldsymbol{\Omega}_o \tilde{\boldsymbol{\theta}}_o). \qquad (17)$$

The new term $\mathbf{p}_o$ appears in (15), (16) due to the introduction of $\mathbf{A}_o$, $\mathbf{B}_o$ in (12)−(14). Under assumption 2 for $\mathbf{y} \in Y$, $\mathbf{v} \in V$ all solutions of the system (13) are bounded, i.e. there exists $k_{\Omega,o} > 0$ such that $|\boldsymbol{\Omega}_o(t)| \leq k_{\Omega,o}$ for all $t \geq 0$. Then $|\mathbf{d}_v| \leq [k_\phi + k_G |\boldsymbol{\theta}| + |\mathbf{L}_o|]|\mathbf{v}|$ and for $\boldsymbol{\theta} \in \Theta$, $\mathbf{v} \in V$ the signal $\mathbf{d}_v$ remains bounded. The signal $\mathbf{p}_o$ is bounded for any $\rho(t) \in \Upsilon$, $\mathbf{x}(t) \in X$, $\mathbf{u}(t) \in U$. Therefore, if assumption 2 is satisfied, the solutions of the system (16) are bounded. In addition, if the signal $\mathbf{C}^T \boldsymbol{\Omega}_o^T(t)$ is persistently exciting, then from lemma 1 the system (17) solutions remain bounded. Since $\boldsymbol{\varepsilon}_o(t) = \boldsymbol{\delta}_o(t) - \boldsymbol{\Omega}_o(t)\boldsymbol{\theta}$ for all $t \geq 0$, the observation error $\boldsymbol{\varepsilon}_o(t)$ is bounded. Therefore, the first part of the theorem is proven, and the solutions of the system (15)−(17) remain bounded provided that $\mathbf{x}(t) \in X$, $\mathbf{u}(t) \in U$, $\mathbf{v}(t) \in V$, $t \geq 0$.

Now, let $\mathbf{v}(t) = 0$ for all $t \geq 0$, that implies $\mathbf{d}_v(t) = 0$, $t \geq 0$. Since $0 \prec \mathbf{G}(\mathbf{y} + \mathbf{v})$ for all $\mathbf{y}(t) \in Y$, $\mathbf{v}(t) \in V$, $t \geq 0$, then monotonicity of the system (13) ensures that $\boldsymbol{\Omega}_o(t) \prec 0$ for all $t \geq 0$ and $o \in \{m, M\}$ for $\boldsymbol{\Omega}_o(0) = 0$. In the equation (14) the gain matrix $\mathbf{\Gamma}_o \mathbf{\Omega}_o^T(t)\mathbf{C}^T\mathbf{C}\mathbf{\Omega}_o(t)$, $t \geq 0$ is positive semidefinite and not negative elementwise for both $o \in \{m, M\}$ due to $0 \prec \mathbf{C}$ (the system (14) is competitive [29]). The matrix coefficients $\mathbf{\Gamma}_o$, $o \in \{m, M\}$ define the rate of changes for the variables $\hat{\mathbf{\theta}}_0$. A modification of $\mathbf{\Gamma}_o$, $o \in \{m, M\}$ does not affect on behavior of the variables $\mathbf{\Omega}_o^T(t)\mathbf{C}^T\mathbf{C}\mathbf{\Omega}_o(t)$ and $\mathbf{\Omega}_o^T(t)\mathbf{C}^T\mathbf{C}\mathbf{\varepsilon}_o(t)$ (they are defined by the decoupled from (14) equations (12), (13) and their initial conditions). If $\mathbf{\Gamma}_o$, $o \in \{m, M\}$ are chosen sufficiently small, then the variables $\hat{\mathbf{\theta}}_0(t)$ become "slowly-varying" in the system (3), (12)−(14) and the variables $\mathbf{\Omega}_o(t)$ and $\mathbf{\varepsilon}_o(t)$ are the "fast" ones. In such conditions, it is possible to apply averaging technique for the equation (14) simplification [5], [27]:

$$\dot{\hat{\mathbf{\theta}}}_o(t) = \mathbf{\Gamma}_o [\mathbf{b}_o - \mathbf{R}_o \hat{\mathbf{\theta}}_o(t)]. \tag{18}$$

The matrices $\mathbf{R}_o$, $o \in \{m, M\}$ are positive definite due to PE condition ($\mathbf{R}_o \geq 0.5\vartheta_o / \ell_o \mathbf{I}_q$ according to lemma A1 from [9]). The system (18) is competitive and stable. The solutions of the system (18) asymptotically converge to the equilibrium $\hat{\mathbf{\theta}}_O^\infty = \mathbf{R}_O^{-1}\mathbf{b}_o$. If $\mathbf{\theta}_M < \mathbf{R}_m^{-1}\mathbf{b}_m$ and $\mathbf{R}_M^{-1}\mathbf{b}_M < \mathbf{\theta}_m$, then using relations between solutions of stable averaged system and the original one (Theorem 5.5.1 in [27]) we get that

$$\lim_{t \to +\infty} \hat{\mathbf{\theta}}_m(t) \geq \mathbf{\theta}_M, \quad \lim_{t \to +\infty} \hat{\mathbf{\theta}}_M(t) \leq \mathbf{\theta}_m.$$

This fact implies that the same relations hold in backward time (for the initial conditions $\tilde{\mathbf{\theta}}_m(0) \leq 0$, $\tilde{\mathbf{\theta}}_M(0) \geq 0$) and $\tilde{\mathbf{\theta}}_m(t) \leq 0$, $\tilde{\mathbf{\theta}}_M(t) \geq 0$ for all $t \geq 0$.

The part (ii).a of the theorem has been proven. The part (ii).b can be proven in the same way. ∎

Theorem 1 establishes the conditions under which the estimation of the set of possible values for $\mathbf{\theta}$ is guaranteed. These conditions restrict admissible values for initial conditions of the system (12)−(14) and the gains $\mathbf{\Gamma}_O$, $o \in \{m, M\}$. For the given set $X$ the conditions $\mathbf{\varepsilon}_m(0) \geq 0$, $\mathbf{\varepsilon}_M(0) \leq 0$ can be easily realized.

The most restrictive condition of the theorem deals with $\mathbf{R}_o$ and $\mathbf{b}_o$ computation for $o \in \{m, M\}$, they can be computed only asymptotically (afterwards the observer (12)−(14) runs). However, these quantities can be used to test reliability of the observers. The values $\hat{\mathbf{\theta}}_O^\infty = \mathbf{R}_O^{-1}\mathbf{b}_o$, $o \in \{m, M\}$ can be evaluated and compared on-line with $\mathbf{\theta}_m$ and $\mathbf{\theta}_M$, i.e. the estimates

$$\hat{\mathbf{b}}_o(t) = -t^{-1}\int_0^t \mathbf{\Omega}_o^T(\tau)\mathbf{C}^T\mathbf{C}\mathbf{\varepsilon}_o(\tau)d\tau, \quad \hat{\mathbf{R}}_o(t) = t^{-1}\int_0^t \mathbf{\Omega}_o^T(\tau)\mathbf{C}^T\mathbf{C}\mathbf{\Omega}_o(\tau)d\tau \tag{19}$$

are well defined for all finite $t \geq \ell_o$, $o \in \{m, M\}$ (by lemma A1 from [9], the matrix $\hat{\mathbf{R}}_o(t)$ is not singular for $t \geq \ell_o$) and the variable $\bar{\mathbf{\theta}}_o^\infty(t) = \hat{\mathbf{R}}_o^{-1}(t)\hat{\mathbf{b}}_o(t)$ can be used for $\hat{\mathbf{\theta}}_o^\infty$ evaluation. Therefore, while the restrictions $\bar{\mathbf{\theta}}_o^\infty(t) \approx \hat{\mathbf{\theta}}_o^\infty$, $o \in \{m, M\}$ required in theorem 1 are satisfied, the observers generate reliable interval estimates for the vector $\mathbf{\theta}$.

From another point of view, theorem 1 fixes initial conditions for the systems (12)−(14), i.e. if the property $\mathbf{x}_m \leq \mathbf{x} \leq \mathbf{x}_M$ holds for all $\mathbf{x} \in X$ for some $\mathbf{x}_m \in R^n$, $\mathbf{x}_M \in R^n$, then the conditions of the part (ii).a of theorem 1 are satisfied taken $\mathbf{\xi}_m(0) = \mathbf{x}_m$, $\mathbf{\xi}_M(0) = \mathbf{x}_M$, $\mathbf{\Omega}_o(0) = 0$, $o \in \{m, M\}$, $\hat{\mathbf{\theta}}_m(0) = \mathbf{\theta}_m$, $\hat{\mathbf{\theta}}_m(0) = \mathbf{\theta}_M$. Therefore, in the system (3), (12)−(14) the unspecified initial conditions are $\mathbf{x}(0) \in X$ only, then $\mathbf{R}_o$ and $\mathbf{b}_o$, $o \in \{m, M\}$ are functions of $\mathbf{x}(0)$ (assuming for simplicity that $\mathbf{v}(t) = 0$). If the system (3) is also monotone, then computation of $\mathbf{R}_o$ and $\mathbf{b}_o$, $o \in \{m, M\}$ for the cases $\mathbf{x}(0) \in \{\mathbf{x}_m, \mathbf{x}_M\}$ with $\mathbf{\theta} \in \{\mathbf{\theta}_m, \mathbf{\theta}_M\}$ has to provide worst-case estimates on the values of $\mathbf{R}_o$ and $\mathbf{b}_o$, $o \in \{m, M\}$.

**Remark 2**. The necessity of $\mathbf{R}_o$, $\mathbf{b}_o$, $o \in \{m, M\}$ computation and the idea of the observers (12)–(14) design can be clarified in other words for the case of assumption 1 ($\mathbf{L}_m = \mathbf{L}_M = \mathbf{L}$), when $\mathbf{x}(t) \geq 0$, $\mathbf{u}(t) \geq 0$ for all $t \geq 0$. In such situation $\mathbf{p}_m(t) \geq 0$, $\mathbf{p}_M(t) \leq 0$. Define $\mathbf{E}_\Omega = \mathbf{\Omega} - \mathbf{\Omega}_o$, where $\mathbf{\Omega}$ is the system (7) solution with $\mathbf{\Omega}(0) = 0$, then

$$\dot{\mathbf{E}}_\Omega = [\mathbf{A}_o - \mathbf{LC}]\mathbf{E}_\Omega + [\mathbf{A}(\rho(t)) - \mathbf{A}_o]\mathbf{\Omega}.$$

The system (7) is stable from assumption 1, cooperative ($\mathbf{A}_m - \mathbf{LC} \prec \mathbf{A}(\rho(t)) - \mathbf{LC} \prec \mathbf{A}_M - \mathbf{LC}$ for all $t \geq 0$ and both $\mathbf{A}_m - \mathbf{LC}$ and $\mathbf{A}_M - \mathbf{LC}$ are cooperative from assumption 2) with negative input and zero initial conditions, therefore, $\mathbf{\Omega}(t) \prec 0$ for all $t \geq 0$ (indeed, $\dot{\mathbf{\Omega}}(0) \leq 0$ and if $\Omega_{i,j}(t)$, $1 \leq i \leq n$, $1 \leq j \leq q$ approaches zero from below, then $\dot{\Omega}_{i,j}(t)$ becomes negative ensuring that $\mathbf{\Omega}(t) \prec 0$ for all $t \geq 0$). Thus, $[\mathbf{A}(\rho(t)) - \mathbf{A}_m]\mathbf{\Omega}(t) \prec 0$ and $0 \prec [\mathbf{A}(\rho(t)) - \mathbf{A}_M]\mathbf{\Omega}(t)$, that under assumption 2 means for $\mathbf{\Omega}_o(0) = 0$:

$$\mathbf{\Omega}_M(t) \prec \mathbf{\Omega}(t) \prec \mathbf{\Omega}_m(t) \prec 0 \text{ for all } t \geq 0.$$

Cooperativeness of the matrix $\mathbf{A}_o - \mathbf{LC}$ in the system (16) implies that $\mathbf{\delta}_m(t) \geq 0$, $\mathbf{\delta}_M(t) \leq 0$ for all $t \geq 0$ provided that $\mathbf{\delta}_m(0) \geq 0$, $\mathbf{\delta}_M(0) \leq 0$ respectively (the conditions $\mathbf{\delta}_m(0) \geq 0$, $\mathbf{\delta}_M(0) \leq 0$ are satisfied for $\mathbf{\varepsilon}_m(0) \geq 0$ and $\mathbf{\varepsilon}_M(0) \leq 0$ since $\mathbf{\Omega}_o(0) = 0$).

Further, in the equation (17) the gain matrix $\mathbf{\Gamma}_o \mathbf{\Omega}_o^T(t)\mathbf{C}^T\mathbf{C}\mathbf{\Omega}_o(t)$, $t \geq 0$ is positive semidefinite and not negative elementwise for both $o \in \{m, M\}$ (the system (17) is competitive [29]), $\mathbf{\Gamma}_m \mathbf{\Omega}_m^T(t)\mathbf{C}^T\mathbf{C}\mathbf{\delta}_m(t) \leq 0$ and $\mathbf{\Gamma}_M \mathbf{\Omega}_M^T(t)\mathbf{C}^T\mathbf{C}\mathbf{\delta}_M(t) \geq 0$ for all $t \geq 0$. If $\mathbf{\Gamma}_o$, $o \in \{m, M\}$ are chosen sufficiently small, then the variables $\tilde{\mathbf{\theta}}_0(t)$ become "slowly-varying" in the system (3), (13), (15)–(17) and the variables $\mathbf{\Omega}_o(t)$ and $\mathbf{\delta}_o(t)$ are the "fast" ones. Under these conditions averaging technique gives:

$$\dot{\tilde{\mathbf{\theta}}}_o(t) = \mathbf{\Gamma}_o[\mathbf{h}_o - \mathbf{R}_o \tilde{\mathbf{\theta}}_o(t)], \ \mathbf{h}_o = \lim_{T \to +\infty} T^{-1} \int_0^T \mathbf{\Omega}_o^T(t)\mathbf{C}^T\mathbf{C}\mathbf{\delta}_o(t)dt. \tag{20}$$

Note, that $\mathbf{\Omega}_o^T(t)\mathbf{C}^T\mathbf{C}\mathbf{\delta}_o(t)$ and $\mathbf{\Omega}_o^T(t)\mathbf{C}^T\mathbf{C}\mathbf{\Omega}_o(t)$ are elementwise sign definite functions, therefore, $\mathbf{h}_o$ and $\mathbf{R}_o$ inherit this property, namely $\mathbf{h}_m \leq 0 \leq \mathbf{h}_M$; $\mathbf{R}_o = \mathbf{R}_o^T > 0$, $0 \prec \mathbf{R}_o$, $o \in \{m, M\}$.

Additionally, since $\mathbf{\Omega}_M(t) \prec \mathbf{\Omega}_m(t) \prec 0$ for all $t \geq 0$ we have $\mathbf{R}_m \prec \mathbf{R}_M$. Thus, the system (20) is competitive and stable. The solutions of the system (20) converge asymptotically to the equilibrium $\tilde{\mathbf{\theta}}_o^\infty = \mathbf{R}_o^{-1}\mathbf{h}_o$. In addition, if $\mathbf{R}_m^{-1}\mathbf{h}_m \leq 0$ and $\mathbf{R}_M^{-1}\mathbf{h}_M \geq 0$, then

$$\lim_{t \to +\infty} \tilde{\mathbf{\theta}}_m(t) \leq 0, \ \lim_{t \to +\infty} \tilde{\mathbf{\theta}}_M(t) \geq 0.$$

For competitive systems this fact implies that $\tilde{\mathbf{\theta}}_m(t) \leq 0$, $\tilde{\mathbf{\theta}}_M(t) \geq 0$ for all $t \geq 0$ for the initial conditions $\tilde{\mathbf{\theta}}_m(0) \leq 0$, $\tilde{\mathbf{\theta}}_M(0) \geq 0$, that is exactly the conclusion of part (ii).a of theorem 1 (the part (ii).b can be illustrated by the case $\mathbf{x}(t) \leq 0$, $\mathbf{u}(t) \leq 0$ for all $t \geq 0$).

Unfortunately, all these nice monotonicity properties for $\mathbf{h}_o$ and $\mathbf{R}_o$, $o \in \{m, M\}$ are not enough to ensure $\mathbf{R}_m^{-1}\mathbf{h}_m \leq 0$ and $\mathbf{R}_M^{-1}\mathbf{h}_M \geq 0$ (the inverse matrices $\mathbf{R}_o^{-1}$ are not elementwise sign definite in general case). As a result, the requirement on $\mathbf{R}_o^{-1}\mathbf{b}_o$ on-line checking is introduced in theorem 1. □

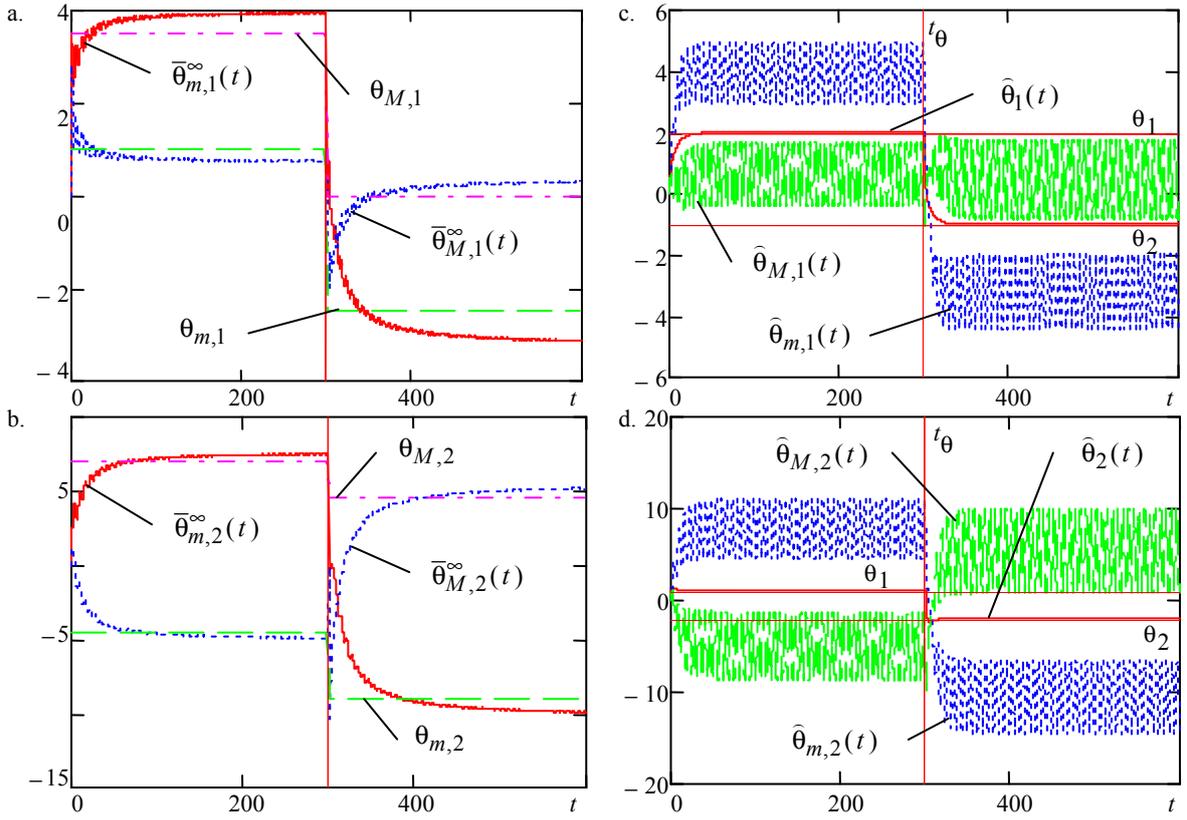

Fig. 1. Results of simulation in example 1 (without disturbances): $\hat{\boldsymbol{\theta}}_o^\infty$ ((a), (b)) and $\hat{\boldsymbol{\theta}}_o$ ((c), (d)), $o \in \{m, M\}$.

**R e m a r k  3**. Let us stress that PE property of the signals $\boldsymbol{\Omega}_o^T(t)\mathbf{C}^T$, $o \in \{m, M\}$ can also be checked on-line by computing the integrals

$$\int_t^{t+\ell_o} \boldsymbol{\Omega}_o^T(\tau)\mathbf{C}^T\mathbf{C}\boldsymbol{\Omega}_o(\tau)d\tau, \ o \in \{m, M\}$$

for some $\ell_o > 0$ for all $t \geq 0$. While these integrals result in a nonsingular matrix, the PE property holds. According to lemma A1 in [9], non-singularity of these integrals are equivalent to the same property of the following integral:

$$t^{-1}\int_0^t \boldsymbol{\Omega}_o^T(\tau)\mathbf{C}^T\mathbf{C}\boldsymbol{\Omega}_o(\tau)d\tau,$$

that coincides with $\hat{\mathbf{R}}_o(t)$ from (19). Thus, by calculating (19), it is possible to check on-line PE properties for $\boldsymbol{\Omega}_o^T(t)\mathbf{C}^T$, $o \in \{m, M\}$, simultaneously with verification of the conditions on $\mathbf{R}_o^{-1}\mathbf{b}_o$, $o \in \{m, M\}$.  □

**R e m a r k  4**. If the functions $\mathbf{C}\boldsymbol{\Omega}_o(t)$ and $\mathbf{C}\boldsymbol{\varepsilon}_o(t)$ are $T$-periodic, then the limits can be dropped in the definitions of $\mathbf{h}_o$ and $\mathbf{R}_o$, $o \in \{m, M\}$ in theorem 1 formulation [27]. In this case, on-line verification of the conditions for $\mathbf{R}_o^{-1}\mathbf{b}_o$ via (19) becomes trivial.  □

Fulfillment of the conditions $\boldsymbol{\theta}_M < \mathbf{R}_m^{-1}\mathbf{b}_m$, $\mathbf{R}_M^{-1}\mathbf{b}_M < \boldsymbol{\theta}_m$ or $\boldsymbol{\theta}_M < \mathbf{R}_M^{-1}\mathbf{b}_M$, $\mathbf{R}_m^{-1}\mathbf{b}_m < \boldsymbol{\theta}_m$ implies that the lower and upper estimates of possible values of $\hat{\boldsymbol{\theta}}_o$, $o \in \{m, M\}$ lie outside of the admissible values interval $[\boldsymbol{\theta}_m, \boldsymbol{\theta}_M]$ for the vector of unknown parameters $\boldsymbol{\theta}$. However, this fact does not mean that the observer (12)−(14) can not improve available *a priori* estimate on the admissible interval $[\boldsymbol{\theta}_m, \boldsymbol{\theta}_M]$. The variables $\hat{\boldsymbol{\theta}}_o$, $o \in \{m, M\}$ converge to these conservative asymptotic estimates $\mathbf{R}_o^{-1}\mathbf{b}_o$ for sufficiently small values of $\boldsymbol{\Gamma}_o$. By closing the gains $\boldsymbol{\Gamma}_o$, $o \in \{m, M\}$ to the boundary $\bar{\Gamma}$ it is possible to compute a more accurate estimate on admissible interval values for $\theta$, that we are going to show in the following example.

**Example 1**. Let

$$\mathbf{A}(t) = \begin{bmatrix} -1+0.5\sin(t) & 1 & 0 \\ 1.2 & -2+0.3\cos(3t) & 1.3 \\ 0 & 1 & -3+0.6\cos(2t) \end{bmatrix}, \quad \mathbf{B} = \begin{bmatrix} 0 \\ 0 \\ 0 \end{bmatrix}, \quad \mathbf{C} = \begin{bmatrix} 1 & 0 & 0 \\ 0 & 1 & 0 \end{bmatrix},$$

$$\mathbf{G}(t) = \begin{bmatrix} 0 & 1 \\ 1-0.2\sin(2t) & 0 \\ 0 & 1+0.3\sin(3t) \end{bmatrix}.$$

In this example, we assume that the exact dependence of the matrix $\mathbf{A}$ on time argument is not known and only majorant matrices are available:

$$\mathbf{A}_m = \begin{bmatrix} -1.5 & 1 & 0 \\ 1.2 & -2.3 & 1.3 \\ 0 & 1 & -3.6 \end{bmatrix}, \quad \mathbf{A}_M = \begin{bmatrix} -0.5 & 1 & 0 \\ 1.2 & -1.7 & 1.3 \\ 0 & 1 & -2.4 \end{bmatrix},$$

while the matrix function $\mathbf{G}(t)$ is measured as it is required in the system (1). Assume that

$$\theta(t) = \begin{cases} \theta_1 & \text{if } 0 \le t \le t_\theta; \\ \theta_2 & \text{if } t_\theta < t \le t_k, \end{cases} \quad \theta_1 = \begin{bmatrix} 2 \\ 1 \end{bmatrix}, \quad \theta_2 = \begin{bmatrix} -1 \\ -2 \end{bmatrix},$$

where $t_f = 600$ is the time of simulation and $t_\theta = 0.5 t_f$. Let

$$\mathbf{L}_m = \mathbf{L}_M = \mathbf{L} = \begin{bmatrix} 2 & 0 & 0 \\ 0 & 3 & 1 \end{bmatrix}^T,$$

then assumption 2 holds for

$$\mathbf{A}_m - \mathbf{LC} = \begin{bmatrix} -3.5 & 1 & 0 \\ 1.2 & -5.3 & 1.3 \\ 0 & 0 & -3.6 \end{bmatrix}, \quad \mathbf{A}_M - \mathbf{LC} = \begin{bmatrix} -2.5 & 1 & 0 \\ 1.2 & -4.7 & 1.3 \\ 0 & 0 & -2.4 \end{bmatrix}$$

and $\theta_m = [1 \; -4.5]^T$, $\theta_M = [3.5 \; 7]^T$ for $0 \le t \le t_\theta$ and $\theta_m = [-2.5 \; -9]^T$, $\theta_M = [0 \; 4.5]^T$ for $t_\theta \le t \le t_f$.

Let $\mathbf{x}(0) = [1\;1\;1]^T$ and $\Gamma = \Gamma_m = \Gamma_M = 5\mathbf{I}_2$. The results of (19) computations and on-line graphical checking the conditions on $\mathbf{R}_o^{-1} \mathbf{b}_o$, $o \in \{m, M\}$ are shown in Fig. 1,a and b. As we can deduce from these figures the conditions of the point (ii).a of theorem 1 are satisfied for $0 \le t \le t_\theta$, and conditions of the point (ii).b are satisfied for $t_\theta \le t \le t_f$. The variables $\hat{\theta}$ (the estimate of the ideal observer (6)–(8)), $\hat{\theta}_m$ and $\hat{\theta}_M$ are plotted in Fig. 1,c and d for the case without disturbances. The variables $\hat{\theta}$, $\hat{\theta}_m$ and $\hat{\theta}_M$ for the case of a stochastic noise presence with $|\mathbf{v}(t)| \le 1$ are shown in Fig. 2. □

Before we continue it is worth to emphasize one feature of the proposed set adaptive observers illustrated by figures 1 and 2. The purpose is not the exact estimation of the values of uncertain parameters, but to evaluate the set or the interval of admissible values for such parameters. Therefore, the lower or upper estimate may have a different sign with respect to the real value of the parameter. The accuracy of the proposed approach is characterized by the interval length comparing with the "size" of uncertainty and complexity presented in the estimated system. In the situation when it is possible to design a conventional observer converging to exact values of state $\mathbf{x}$ or parameters $\mathbf{d}$ there is no need in interval observation. However, frequently for complex nonlinear systems with signal and parametric uncertainties the design of conventional exact observers is not possible. In this case the interval observation becomes useful, being the only available solution in practice.

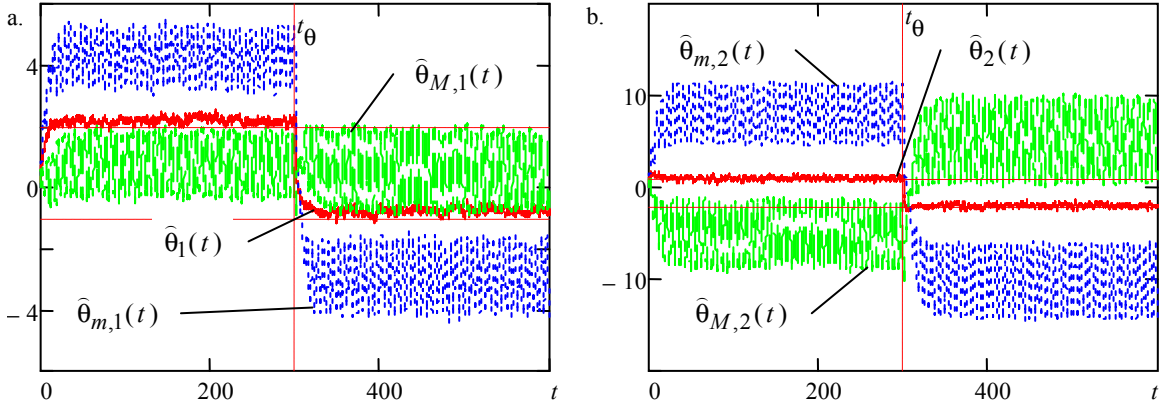

Fig. 2. Results of simulation in example 1 (with disturbances): $\hat{\boldsymbol{\theta}}_o$, $o \in \{m, M\}$.

*D. Cooperative case*

Competitiveness of the adaptive observers (12)−(14) follows by assumption that $0 \prec \mathbf{C}$. Such restriction is natural and corresponds to situation when some part of the state space vector $\mathbf{x}$ coordinates is available for measurements. Relaxation of this assumption leads to the case when the matrices $-\boldsymbol{\Gamma}_o \boldsymbol{\Omega}_o^T(t) \mathbf{C}^T \mathbf{C} \boldsymbol{\Omega}_o(t)$, $o \in \{m, M\}$ may become cooperative.

**Theorem 2**. *Let assumption 2 hold, and $\mathbf{x}(t) \in X$, $\mathbf{u}(t) \in U$, $\mathbf{v}(t) \in V$, $\rho(t) \in \Upsilon$ and $\boldsymbol{\theta} \in \Theta$ for all $t \geq 0$, and assume that the signals $\boldsymbol{\Omega}_o^T(t) \mathbf{C}^T$ are $(\ell_o, \vartheta_o)$–PE for some $\ell_o > 0$, $\vartheta_o > 0$, $o \in \{m, M\}$. Then*

(i) *for all $t \in R$ and $o \in \{m, M\}$ the solutions $\boldsymbol{\zeta}_o(t)$, $\boldsymbol{\Omega}_o(t)$ and $\hat{\boldsymbol{\theta}}_o(t)$ of the system (12)−(14) are bounded provided that $\mathbf{v}(t) \in V$, $t \geq 0$;*

(ii) *let $\mathbf{v}(t) \equiv 0$ and the matrices $-\boldsymbol{\Gamma}_o \boldsymbol{\Omega}_o^T(t) \mathbf{C}^T \mathbf{C} \boldsymbol{\Omega}_o(t)$ be cooperative for all $t \geq 0$, $o \in \{m, M\}$,*

    a. *if for all $t \geq 0$ and $o \in \{m, M\}$, $O = \{m, M\} \setminus o$,*

$$\boldsymbol{\Gamma}_o \boldsymbol{\Omega}_o^T(t) \mathbf{C}^T \mathbf{C}[\boldsymbol{\varepsilon}_o(t) + \boldsymbol{\Omega}_o(t) \boldsymbol{\theta}_o] \geq 0, \; \boldsymbol{\Gamma}_o \boldsymbol{\Omega}_o^T(t) \mathbf{C}^T \mathbf{C} \boldsymbol{\Omega}_o(t)(\boldsymbol{\theta}_O - \boldsymbol{\theta}_o) \geq 0;$$

$$\boldsymbol{\Gamma}_O \boldsymbol{\Omega}_O^T(t) \mathbf{C}^T \mathbf{C}[\boldsymbol{\varepsilon}_O(t) + \boldsymbol{\Omega}_O(t) \boldsymbol{\theta}_O] \leq 0, \; \boldsymbol{\Gamma}_O \boldsymbol{\Omega}_O^T(t) \mathbf{C}^T \mathbf{C} \boldsymbol{\Omega}_O(t)(\boldsymbol{\theta}_o - \boldsymbol{\theta}_O) \leq 0,$$

*then $\hat{\boldsymbol{\theta}}_o(t) \leq \boldsymbol{\theta} \leq \hat{\boldsymbol{\theta}}_O(t)$, $t \geq 0$.*

    b. *there exists a matrix $\bar{\boldsymbol{\Gamma}}$ such that for all $0 \prec \boldsymbol{\Gamma}_o \prec \bar{\boldsymbol{\Gamma}}$, $o \in \{m, M\}$ if the signals $\mathbf{C}\boldsymbol{\varepsilon}_o(t)$ and $\mathbf{C}\boldsymbol{\Omega}_o(t)$ are $T$-periodical for some $T > 0$, $t \geq 0$ and for all $t \geq 0$ and $o \in \{m, M\}$, $O = \{m, M\} \setminus o$,*

$$\mathbf{b}_o \leq \mathbf{R}_o \boldsymbol{\theta}_o, \; \mathbf{R}_o(\boldsymbol{\theta}_O - \boldsymbol{\theta}_o) \geq 0; \; \mathbf{b}_O \geq \mathbf{R}_O \boldsymbol{\theta}_O, \; \mathbf{R}_O(\boldsymbol{\theta}_o - \boldsymbol{\theta}_O) \leq 0,$$

*then $\hat{\boldsymbol{\theta}}_o(t) \leq \boldsymbol{\theta} \leq \hat{\boldsymbol{\theta}}_O(t)$, $t \geq 0$, where*

$$\mathbf{b}_o = -T^{-1} \int_0^T \boldsymbol{\Omega}_o^T(\tau) \mathbf{C}^T \mathbf{C} \boldsymbol{\varepsilon}_o(\tau) d\tau, \; \mathbf{R}_o = T^{-1} \int_0^T \boldsymbol{\Omega}_o^T(\tau) \mathbf{C}^T \mathbf{C} \boldsymbol{\Omega}_o(\tau) d\tau.$$

**Proof**. The part (i) of the theorem can be proven in the same way as in theorem 1. Under conditions of the part (ii).a the system (14) is asymptotically stable cooperative with sign definite inputs. Rewriting the system (14) equations we obtain:

$$\dot{\tilde{\boldsymbol{\theta}}}_o = -\boldsymbol{\Gamma}_o \boldsymbol{\Omega}_o^T \mathbf{C}^T \mathbf{C} \boldsymbol{\varepsilon}_o - \boldsymbol{\Gamma}_o \boldsymbol{\Omega}_o^T \mathbf{C}^T \mathbf{C} \boldsymbol{\Omega}_o \tilde{\boldsymbol{\theta}}_o - \boldsymbol{\Gamma}_o \boldsymbol{\Omega}_o^T \mathbf{C}^T \mathbf{C} \boldsymbol{\Omega}_o \boldsymbol{\theta}, \; \tilde{\boldsymbol{\theta}}_o = \hat{\boldsymbol{\theta}}_o - \boldsymbol{\theta}, \; o \in \{m, M\}. \quad (21)$$

The matrices $-\mathbf{\Gamma}_o \mathbf{\Omega}_o^T(t)\mathbf{C}^T\mathbf{C}\mathbf{\Omega}_o(t)$, $o \in \{m, M\}$ are cooperative and stable (persistency of excitation ensures the last property). If the signals $-\mathbf{\Gamma}_o \mathbf{\Omega}_o^T \mathbf{C}^T \mathbf{C} \boldsymbol{\delta}_o = -\mathbf{\Gamma}_o \mathbf{\Omega}_o^T \mathbf{C}^T \mathbf{C} \boldsymbol{\varepsilon}_o - \mathbf{\Gamma}_o \mathbf{\Omega}_o^T \mathbf{C}^T \mathbf{C} \mathbf{\Omega}_o \boldsymbol{\theta}$, $o \in \{m, M\}$ are sign definite, then applying monotonicity, it is possible to substantiate the desired relations between $\hat{\boldsymbol{\theta}}_m(t)$, $\hat{\boldsymbol{\theta}}_M(t)$ and $\boldsymbol{\theta}$. Let us evaluate the signal $-\mathbf{\Gamma}_o \mathbf{\Omega}_o^T \mathbf{C}^T \mathbf{C} \boldsymbol{\delta}_o$ sign using the given measurable information. Note that

$$-\mathbf{\Gamma}_o \mathbf{\Omega}_o^T \mathbf{C}^T \mathbf{C} \boldsymbol{\delta}_o = -\mathbf{\Gamma}_o \mathbf{\Omega}_o^T \mathbf{C}^T \mathbf{C} \boldsymbol{\varepsilon}_o - \mathbf{\Gamma}_o \mathbf{\Omega}_o^T \mathbf{C}^T \mathbf{C} \mathbf{\Omega}_o \boldsymbol{\theta}_o - \mathbf{\Gamma}_o \mathbf{\Omega}_o^T \mathbf{C}^T \mathbf{C} \mathbf{\Omega}_o (\boldsymbol{\theta} - \boldsymbol{\theta}_o),$$

and the sign of the signals $-\mathbf{\Gamma}_o \mathbf{\Omega}_o^T \mathbf{C}^T \mathbf{C} \boldsymbol{\varepsilon}_o - \mathbf{\Gamma}_o \mathbf{\Omega}_o^T \mathbf{C}^T \mathbf{C} \mathbf{\Omega}_o \boldsymbol{\theta}_o$, $o \in \{m, M\}$ can be verified on-line. The sign of the last term for all $\boldsymbol{\theta}_m \le \boldsymbol{\theta} \le \boldsymbol{\theta}_M$ lies between zero and the sign of $\mathbf{\Gamma}_o \mathbf{\Omega}_o^T \mathbf{C}^T \mathbf{C} \mathbf{\Omega}_o (\boldsymbol{\theta}_O - \boldsymbol{\theta}_o)$, $o \in \{m, M\}$, $O = \{m, M\} \setminus o$ [1] (the matrix $\mathbf{\Gamma}_o \mathbf{\Omega}_o^T \mathbf{C}^T \mathbf{C} \mathbf{\Omega}_o$ is competitive/monotone). Therefore, the set of implications hold:

$$-\mathbf{\Gamma}_o \mathbf{\Omega}_o^T(t)\mathbf{C}^T \mathbf{C} \boldsymbol{\varepsilon}_o(t) - \mathbf{\Gamma}_o \mathbf{\Omega}_o^T(t)\mathbf{C}^T \mathbf{C} \mathbf{\Omega}_o(t) \boldsymbol{\theta}_o \le 0,\ -\mathbf{\Gamma}_o \mathbf{\Omega}_o^T(t)\mathbf{C}^T \mathbf{C} \mathbf{\Omega}_o(t)(\boldsymbol{\theta}_O - \boldsymbol{\theta}_o) \le 0,\ t \ge 0 \Rightarrow \hat{\boldsymbol{\theta}}_o(t) \le \boldsymbol{\theta};$$

$$-\mathbf{\Gamma}_o \mathbf{\Omega}_o^T(t)\mathbf{C}^T \mathbf{C} \boldsymbol{\varepsilon}_o(t) - \mathbf{\Gamma}_o \mathbf{\Omega}_o^T(t)\mathbf{C}^T \mathbf{C} \mathbf{\Omega}_o(t) \boldsymbol{\theta}_o \ge 0,\ -\mathbf{\Gamma}_o \mathbf{\Omega}_o^T(t)\mathbf{C}^T \mathbf{C} \mathbf{\Omega}_o(t)(\boldsymbol{\theta}_O - \boldsymbol{\theta}_o) \ge 0,\ t \ge 0 \Rightarrow \hat{\boldsymbol{\theta}}_o(t) \ge \boldsymbol{\theta},$$

that implies the theorem claim (ii).a. To prove part (ii).b, assume that norm of the matrices $\mathbf{\Gamma}_o$, $o \in \{m, M\}$ are chosen small enough to ensure that the variables $\hat{\boldsymbol{\theta}}_o(t)$ are slowly-varying in the system (12)−(14). Applying averaging technique for the equation (21) with $T$-periodical right hand side [5], [27] we obtain:

$$\dot{\tilde{\boldsymbol{\theta}}}_o = \mathbf{b}_o - \mathbf{R}_o \tilde{\boldsymbol{\theta}}_o - \mathbf{R}_o \boldsymbol{\theta},\ o \in \{m, M\},$$

where the matrices $\mathbf{R}_o$, $o \in \{m, M\}$ are cooperative and Hurwitz by the same arguments. Again

$$\mathbf{b}_o - \mathbf{R}_o \boldsymbol{\theta} = \mathbf{b}_o - \mathbf{R}_o \boldsymbol{\theta}_o - \mathbf{R}_o (\boldsymbol{\theta} - \boldsymbol{\theta}_o)$$

and the sign of $\mathbf{b}_o - \mathbf{R}_o \boldsymbol{\theta}_o$ can be verified during or before the observers operation and $\mathbf{R}_o (\boldsymbol{\theta} - \boldsymbol{\theta}_o) \in [0, \mathbf{R}_o (\boldsymbol{\theta}_O - \boldsymbol{\theta}_o)]$ for all $\boldsymbol{\theta}_m \le \boldsymbol{\theta} \le \boldsymbol{\theta}_M$ and $o \in \{m, M\}$, $O = \{m, M\} \setminus o$. ∎

The cooperative case is more sophisticated and it requires an on-line verification of a bigger number of conditions. To check constraints imposed on $\mathbf{b}_o$, $\mathbf{R}_o$, $o \in \{m, M\}$ for the system (3) solutions being $T$-periodical asymptotically, the following variables can be computed for $t > T$:

$$\hat{\mathbf{b}}_o(t) = -T^{-1}\int_{t-T}^{t} \mathbf{\Omega}_o^T(\tau)\mathbf{C}^T \mathbf{C} \boldsymbol{\varepsilon}_o(\tau) d\tau,\ \hat{\mathbf{R}}_o(t) = T^{-1}\int_{t-T}^{t} \mathbf{\Omega}_o^T(\tau)\mathbf{C}^T \mathbf{C} \mathbf{\Omega}_o(\tau) d\tau.$$

**E x a m p l e  2**. Let

$$\mathbf{A}(t) = \begin{bmatrix} -1 + 0.1\sin(3t) & 1 & 0.4 + 0.2\sin(3t) \\ 0 & -1 + 0.3\cos(t) & 1 \\ 0.5 + 0.1\cos(2t) & 1 & -2 + 0.2\cos(2t) \end{bmatrix},\ \mathbf{B} = \begin{bmatrix} 0 \\ 0 \\ 0 \end{bmatrix},\ \mathbf{C} = \begin{bmatrix} 1 & 0 & -1 \\ 1 & 1 & 0 \end{bmatrix},$$

$$\mathbf{G}(t) = \begin{bmatrix} 1 & 0 \\ 0.3 + 0.3\sin(2t) & 0 \\ 0 & 0.3 + 0.2\sin(3t) \end{bmatrix}.$$

Again, in this example we assume that the exact dependence of the matrix $\mathbf{A}$ on time argument is not known and only majorant matrices are available:

$$\mathbf{A}_m = \begin{bmatrix} -0.9 & 1 & .6 \\ 0 & -0.7 & 1 \\ 0.6 & 1 & -1.8 \end{bmatrix},\ \mathbf{A}_M = \begin{bmatrix} -1.1 & 1 & 0.2 \\ 0 & -1.3 & 1 \\ 0.4 & 1 & -2.2 \end{bmatrix},$$

while the matrix function $\mathbf{G}(t)$ is measured. Assume that

$$\boldsymbol{\theta}(t) = \begin{cases} \boldsymbol{\theta}_1 & \text{if } 0 \leq t \leq t_\theta; \\ \boldsymbol{\theta}_2 & \text{if } t_\theta < t \leq t_k, \end{cases} \quad \boldsymbol{\theta}_1 = \begin{bmatrix} -.5 \\ -1 \end{bmatrix}, \quad \boldsymbol{\theta}_2 = \begin{bmatrix} 0 \\ -2 \end{bmatrix},$$

where $t_f = 600$ is the time of simulation and $t_\theta = 0.5 t_f$. Let

$$\mathbf{L}_m = \begin{bmatrix} 0 & -1 & 0 \\ 0.5 & 1 & -1 \end{bmatrix}^T, \quad \mathbf{L}_M = \begin{bmatrix} 0 & -1 & 0 \\ 1 & 1 & 0.6 \end{bmatrix}^T,$$

then assumption 2 holds for $\boldsymbol{\theta}_m = [-1 \ -2.5]^T$, $\boldsymbol{\theta}_M = [0.5 \ 0]^T$ and

$$\mathbf{A}_m - \mathbf{L}_m \mathbf{C} = \begin{bmatrix} -1.6 & 0.5 & 0.2 \\ 0 & -2.3 & 0 \\ 1.4 & 2 & -2.2 \end{bmatrix}, \quad \mathbf{A}_M - \mathbf{L}_M \mathbf{C} = \begin{bmatrix} -1.9 & 0 & 0.6 \\ 0 & -1.7 & 0 \\ 0 & 0.4 & -1.8 \end{bmatrix}.$$

---
[1] The symbol \ is used for the set complement.

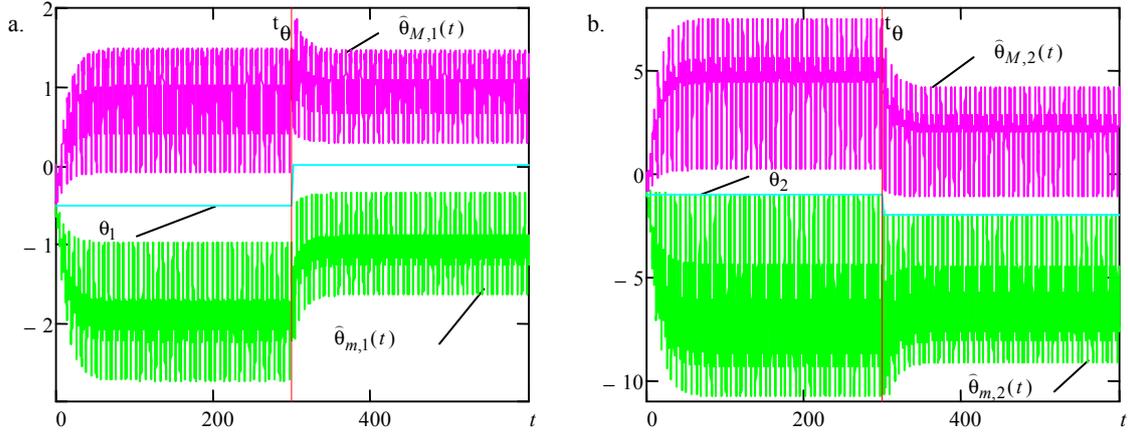

Fig. 3. Results of simulation in example 2 (without disturbances): $\widehat{\boldsymbol{\theta}}_o$, $o \in \{m, M\}$.

Let $\mathbf{x}(0) = [0\ 0\ 0]^T$ and $\boldsymbol{\Gamma} = \boldsymbol{\Gamma}_m = \boldsymbol{\Gamma}_M = diag([40\ 180]^T)$. From the system equations we conclude that the solutions become asymptotically $2\pi$-periodical functions of time. Numerical calculations show that $\mathbf{G}(t)$ is persistently excited with $\ell = 2\pi$, therefore the signals $\boldsymbol{\Omega}_o^T(t) \mathbf{C}^T$, $o \in \{m, M\}$ possess the same property. Numerical calculation of the matrices $-\boldsymbol{\Gamma}_o \boldsymbol{\Omega}_o^T(t) \mathbf{C}^T \mathbf{C} \boldsymbol{\Omega}_o(t)$, $\widehat{\mathbf{b}}_o(t)$, $\widehat{\mathbf{R}}_o(t)$ for both $o \in \{m, M\}$ shows that the conditions

$$\widehat{\mathbf{b}}_m(t) \leq \widehat{\mathbf{R}}_m(t) \boldsymbol{\theta}_m, \ \widehat{\mathbf{R}}_m(t)(\boldsymbol{\theta}_M - \boldsymbol{\theta}_m) \geq 0; \ \widehat{\mathbf{b}}_M(t) \geq \widehat{\mathbf{R}}_M(t) \boldsymbol{\theta}_M, \ \widehat{\mathbf{R}}_M(t)(\boldsymbol{\theta}_m - \boldsymbol{\theta}_M) \leq 0$$

are satisfied for all $t \geq 25$ (the first 25 seconds is the interval of the observer convergence from the chosen zero initial conditions). Therefore, all conditions of theorem 2, part (ii).b hold and it should be $\widehat{\boldsymbol{\theta}}_m(t) \leq \boldsymbol{\theta} \leq \widehat{\boldsymbol{\theta}}_M(t)$, $t \geq 25$, that is confirmed by results of the system simulation presented in Fig. 3. The variables $\widehat{\boldsymbol{\theta}}_m$ and $\widehat{\boldsymbol{\theta}}_M$ for the case of a stochastic noise presence with $|\mathbf{v}(t)| \leq 0.5$ are plotted in Fig. 4. □

**R e m a r k  5**. It is important to note that the conditions of assumption 2 used in theorems 1,2 to substantiate properties of the adaptive set observers are less restrictive than the corresponding conditions of assumption 1 applicable to the conventional adaptive observers (it is hard to compute the matrices $\mathbf{L}$ and $\mathbf{P}$ from assumption 1 in general case). This fact justifies that the set observers can be applied in case where conventional observers can not be realized due to lack of information about the system or plant models complexity. □

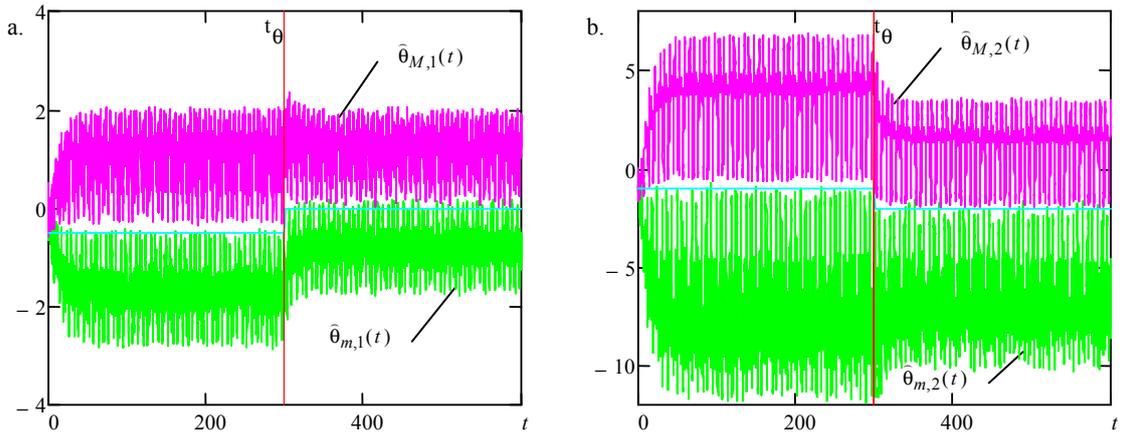

Fig. 4. Results of simulation in example 2 (with disturbances): $\widehat{\boldsymbol{\theta}}_o$, $o \in \{m, M\}$.

## 5. Set state observer

Consider the following observers

$$\dot{\boldsymbol{\xi}}_o = \mathbf{A}_o \boldsymbol{\xi}_o + \mathbf{B}_o \mathbf{u} + \phi(\mathbf{y}_v) + \mathbf{G}(\mathbf{y}_v)\widehat{\boldsymbol{\theta}}_{O_o} + \mathbf{L}_o(\mathbf{y}_v - \mathbf{C}\boldsymbol{\xi}_o), \quad o, O_o \in \{m, M\}, \tag{22}$$

where $\widehat{\boldsymbol{\theta}}_{O_o}$, $O_o \in \{m, M\}$ are generated by (14) and $\boldsymbol{\xi}_o \in R^n$, $o \in \{m, M\}$ are the state estimates. The equation (22) partly repeats (12), however, the state $\boldsymbol{\zeta}_o$, $o \in \{m, M\}$ of the system (12) can not be used for the state $\mathbf{x}$ interval estimation since one of the inequalities $\widehat{\boldsymbol{\theta}}_m < \widehat{\boldsymbol{\theta}}_M$ or $\widehat{\boldsymbol{\theta}}_M < \widehat{\boldsymbol{\theta}}_m$ holds depending on the auxiliary conditions formulated in theorems 1,2. This is why an additional index $O_o$ is introduced in (22). Under conditions of theorems 1,2 the state interval observation via (22) follows by standard arguments [29].

**T h e o r e m  3**. *Let assumption 2 hold, and* $\mathbf{x}(t) \in X$, $\mathbf{u}(t) \in U$, $\mathbf{v}(t) \in V$, $\rho(t) \in \Upsilon$ *and* $\boldsymbol{\theta} \in \Theta$ *for all* $t \geq 0$, *and assume that the signals* $\boldsymbol{\Omega}_o^T(t)\mathbf{C}^T$ *are* $(\ell_o, \vartheta_o)$–*PE for some* $\ell_o > 0$, $\vartheta_o > 0$, $o \in \{m, M\}$. *Then*

(i) *for all* $t \geq 0$ *and* $o \in \{m, M\}$ *the solutions* $\boldsymbol{\xi}_o(t)$, $\boldsymbol{\zeta}_o(t)$, $\boldsymbol{\Omega}_o(t)$ *and* $\widehat{\boldsymbol{\theta}}_o(t)$ *of the system (12)−(14), (22) are bounded provided that* $\mathbf{v}(t) \in V$, $t \geq 0$;

(ii) *let* $\mathbf{v}(t) \equiv 0$, $\mathbf{x}(t) \geq 0$, $\mathbf{u}(t) \geq 0$ *for all* $t \geq 0$ *and theorem 1, part (ii) or theorem 2, part (ii) conditions are verified indicating that* $\widehat{\boldsymbol{\theta}}_o(t) \leq \boldsymbol{\theta} \leq \widehat{\boldsymbol{\theta}}_O(t)$, $o, O \in \{m, M\}$, $t \geq 0$, *then also* $\boldsymbol{\xi}_m(t) \leq \mathbf{x}(t) \leq \boldsymbol{\xi}_M(t)$ *for all* $t \geq 0$ *provided that* $\boldsymbol{\xi}_m(0) \leq \mathbf{x}(0) \leq \boldsymbol{\xi}_M(0)$ *and* $O_m = o$, $O_M = O$ *in (22)*;

(iii) *let* $\mathbf{v}(t) \equiv 0$, $\mathbf{x}(t) \leq 0$, $\mathbf{u}(t) \leq 0$ *for all* $t \geq 0$ *and theorem 1, part (ii) or theorem 2, part (ii) conditions are verified indicating that* $\widehat{\boldsymbol{\theta}}_o(t) \leq \boldsymbol{\theta} \leq \widehat{\boldsymbol{\theta}}_O(t)$, $o, O \in \{m, M\}$, $t \geq 0$, *then also* $\boldsymbol{\xi}_M(t) \leq \mathbf{x}(t) \leq \boldsymbol{\xi}_m(t)$ *for all* $t \geq 0$ *provided that* $\boldsymbol{\xi}_M(0) \leq \mathbf{x}(0) \leq \boldsymbol{\xi}_m(0)$ *and* $O_m = O$, $O_M = o$ *in (22)*.

**P r o o f**. Consider the estimation errors $\mathbf{e}_o = \mathbf{x} - \boldsymbol{\xi}_o$, $o, O_o \in \{m, M\}$,

$$\dot{\mathbf{e}}_o = [\mathbf{A}_o - \mathbf{L}_o\mathbf{C}]\mathbf{e}_o + \mathbf{G}(\mathbf{y}_v)[\boldsymbol{\theta} - \widehat{\boldsymbol{\theta}}_{O_o}] + \mathbf{d}_v + \mathbf{p}_o, \tag{23}$$

$$\mathbf{p}_o = [\mathbf{A}(\rho(t)) - \mathbf{A}_o]\mathbf{x} + [\mathbf{B}(\rho(t)) - \mathbf{B}_o]\mathbf{u}, \quad \mathbf{d}_v = \phi(\mathbf{y}) - \phi(\mathbf{y}_v) + [\mathbf{G}(\mathbf{y}) - \mathbf{G}(\mathbf{y}_v)]\boldsymbol{\theta} - \mathbf{L}\mathbf{v}.$$

Since all conditions of theorem 1, part (i) or theorem 2, part (i) are satisfied, then the solutions $\boldsymbol{\zeta}_o(t)$, $\boldsymbol{\Omega}_o(t)$ and $\widehat{\boldsymbol{\theta}}_o(t)$ are bounded for both $o \in \{m, M\}$. While $\mathbf{x}(t) \in X$, $\mathbf{u}(t) \in U$, $\mathbf{v}(t) \in V$, $\rho(t) \in \Upsilon$ and $\boldsymbol{\theta} \in \Theta$ the signals $\mathbf{p}_o(t)$, $o \in \{m, M\}$ and $\mathbf{d}_v(t)$ stay bounded, and under assumption 2, (23) is an asymptotically stable cooperative linear system with bounded input $\mathbf{G}(\mathbf{y}_v)[\boldsymbol{\theta}-\hat{\boldsymbol{\theta}}_{O_o}]+\mathbf{d}_v+\mathbf{p}_o$, that implies boundedness of the variables $\boldsymbol{\xi}_o(t)$, $o \in \{m, M\}$. The part (i) has been proven.

To substantiate the part (ii) note that in this case $\mathbf{p}_m(t) \geq 0$, $\mathbf{p}_M(t) \leq 0$, $\mathbf{d}_v(t) = 0$ for $t \geq 0$. Then the system (23) with $o = m$ is cooperative with positive input $\mathbf{G}(\mathbf{y})[\boldsymbol{\theta}-\hat{\boldsymbol{\theta}}_o]+\mathbf{p}_m$, by standard arguments in this case, if $\mathbf{e}_m(0) \geq 0$, then the property $\mathbf{e}_m(t) \geq 0$ is preserved for all $t \geq 0$. For $o = M$ the system (23) is cooperative with negative valued input $\mathbf{G}(\mathbf{y})[\boldsymbol{\theta}-\hat{\boldsymbol{\theta}}_O]+\mathbf{p}_M$, that for $\mathbf{e}_M(0) \leq 0$ implies $\mathbf{e}_M(t) \leq 0$, $t \geq 0$. In the case of part (iii), $\mathbf{p}_M(t) \geq 0$, $\mathbf{p}_m(t) \leq 0$, $\mathbf{d}_v(t) = 0$ for all $t \geq 0$. Then the input $\mathbf{G}(\mathbf{y})[\boldsymbol{\theta}-\hat{\boldsymbol{\theta}}_O]+\mathbf{p}_m$ is negative and the input $\mathbf{G}(\mathbf{y})[\boldsymbol{\theta}-\hat{\boldsymbol{\theta}}_o]+\mathbf{p}_M$ is positive, that implies the theorem claim. ∎

For easy reference, the computational procedure is summarized as follows:

1. Take the given sets $X$, $U$, $V$, $Y$, $\Theta$, $\Upsilon$ and compute the bounds $\mathbf{x}_m$, $\mathbf{x}_M$, $\boldsymbol{\theta}_m$ and $\boldsymbol{\theta}_M$.
2. Transform the system (1) to the LPV form (3).
3. Find the matrices $\mathbf{L}_o$, $o \in \{m, M\}$ and verify Assumption 2.
4. Build the set adaptive observer (12)−(14). Calculate (19) and check the PE condition. Distinguish competitive or cooperative cases:
    a. Competitive case ($0 \prec \mathbf{C}$). Verify the properties of either $\overline{\boldsymbol{\theta}}_o^\infty$ or $\hat{\boldsymbol{\theta}}_o^\infty$, $o \in \{m, M\}$ in accordance with the part (ii) of Theorem 1.
    b. Cooperative case (the matrix $-\boldsymbol{\Gamma}_o \boldsymbol{\Omega}_o^T(t) \mathbf{C}^T \mathbf{C} \boldsymbol{\Omega}_o(t)$, $t \geq 0$ is cooperative). Check the inequalities of the part (ii) of Theorem 2.
5. Augment the set state observer (22) and check the conditions of the parts (ii) or (iii) of Theorem 3.

**E x a m p l e  2** (continue). It was shown previously that in this case $\hat{\boldsymbol{\theta}}_m(t) \leq \boldsymbol{\theta} \leq \hat{\boldsymbol{\theta}}_M(t)$ for all $t \geq 25$. Since $\mathbf{u}(t) = 0$ and $\boldsymbol{\theta} \leq 0$, then $\mathbf{x}(t) \leq 0$ for all $t \geq 0$ and the conditions of theorem 3, part (iii) are satisfied. The corresponding trajectory of the state observers (22) is shown in Fig. 5 for two time windows (after and before $t_\theta$). □

As in the ideal case (6)−(8) if some components of $\boldsymbol{\rho}$ are available for measurement (the output $\mathbf{y}$, for instance), then they can be preserved in the matrices $\mathbf{A}_m$, $\mathbf{A}_M$ that become matrix functions $\mathbf{A}_m(\mathbf{y})$, $\mathbf{A}_M(\mathbf{y})$, this idea is illustrated in the next example.

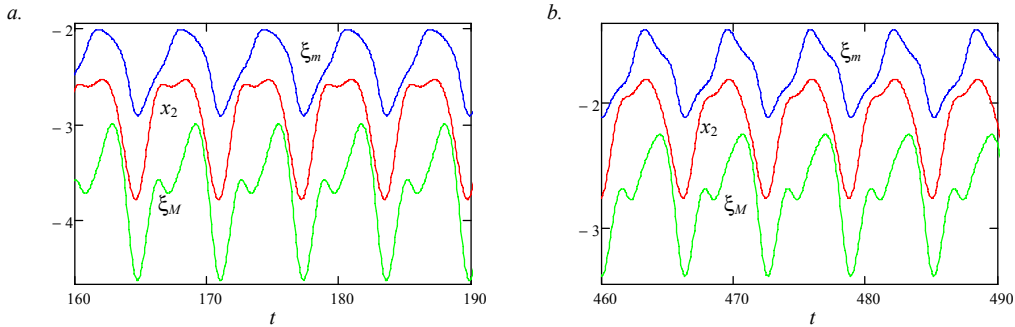

Fig. 5. Results of state estimation in example 2 (without disturbances).

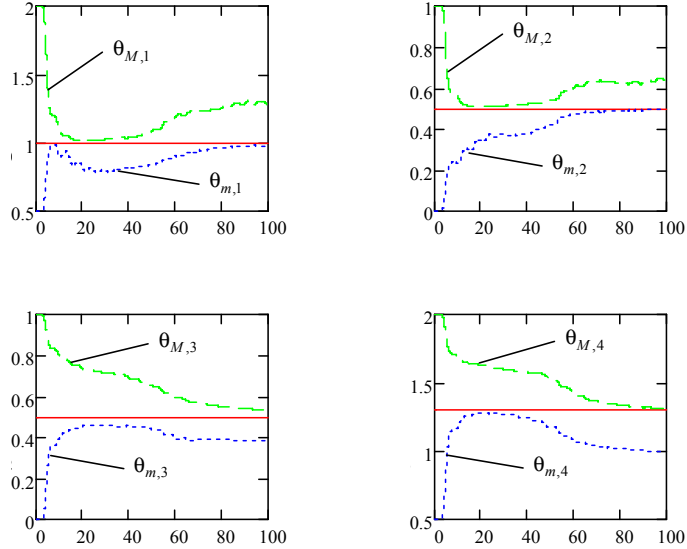

Fig. 6. The parameters set estimation for (24): $\hat{\boldsymbol{\theta}}_o$, $o \in \{m, M\}$.

**E x a m p l e   3**. Consider a double mass model for a vibration crusher [10], the masses correspond to two platforms connecting by springs and excited by rotating motors each. We assume, that movements of platforms are possible in vertical plane only. Mathematical model of the system has form

$$\dot{x}_1 = x_2; \; y_1 = x_1 + v_1;$$
$$\dot{x}_2 = -\beta_1/m(t)x_2 - c/m(t)(x_1 - x_3) - c_0/m(t)x_1 + \theta_1 u_1(t) + \theta_2 u_2(t); \quad (24a)$$

$$\dot{x}_3 = x_4; \; y_2 = x_3 + v_2;$$
$$\dot{x}_4 = -\beta_2/M(t)x_4 + c/M(t)(x_1 - x_3) - c_1/M(t)x_3 + \theta_3 u_1(t) + \theta_4 u_2(t), \quad (24b)$$

where $x_1 \in R$, $x_3 \in R$ are displacements of the platforms from their steady state positions, $\dot{x}_1 \in R$, $\dot{x}_3 \in R$ are velocities of the platforms; $y_1 \in R$, $y_2 \in R$ are noisy measurements; $u_1$, $u_2$ are exciting forces formed by the rotating motors located on the platforms; $\beta_1$, $\beta_2$ are small known friction coefficients; values of spring stickiness $c_1$, $c_0$ are known, the value $c$ of coupling stickiness is unknown; $\boldsymbol{\theta} \in R^4$ is the vector of unknown control gains. Values of masses $m$ and $M$ are assumed unknown and time-varying. Uppers bounds are given for all unknown parameters and the state $\mathbf{x}$: $c_m \leq c \leq c_M$, $m_m \leq m(t) \leq m_M$, $m_m \leq M(t) \leq m_M$, $\boldsymbol{\theta}_m \leq \boldsymbol{\theta} \leq \boldsymbol{\theta}_M$, $\mathbf{x}_m \leq \mathbf{x} \leq \mathbf{x}_M$. The controls are the positive half-period square pulses with amplitude 1 and periods 5 and 6 respectively. Take

$$\mathbf{A}_m = \begin{bmatrix} 0 & 1 & 0 & 0 \\ -(\beta_1 + c_M)m_m^{-1} & -c_0 m_m^{-1} & c_m m_M^{-1} & 0 \\ 0 & 0 & 0 & 1 \\ c_m m_M^{-1} & 0 & -(\beta_2 + c_M)m_m^{-1} & -c_0 m_m^{-1} \end{bmatrix}, \; \mathbf{A}_M = \begin{bmatrix} 0 & 1 & 0 & 0 \\ -(\beta_1 + c_m)m_M^{-1} & -c_0 m_M^{-1} & c_M m_m^{-1} & 0 \\ 0 & 0 & 0 & 1 \\ c_M mm & 0 & -(\beta_2 + c_m)m_M^{-1} & -c_0 m_M^{-1} \end{bmatrix},$$

$$\mathbf{G}(t) = \begin{bmatrix} 0 & 0 & 0 & 0 \\ u_1(t) & u_2(t) & 0 & 0 \\ 0 & 0 & 0 & 0 \\ 0 & 0 & u_1(t) & u_2(t) \end{bmatrix}, \; \mathbf{L}_m = \begin{bmatrix} 1 & 0 \\ -(\beta_1 + c_M)m_m^{-1} & 0 \\ 0 & 1 \\ 0 & -(\beta_2 + c_M)m_m^{-1} \end{bmatrix}, \; \mathbf{L}_M = \begin{bmatrix} 1 & 0 \\ -(\beta_1 + c_m)m_M^{-1} & 0 \\ 0 & 1 \\ 0 & -(\beta_2 + c_m)m_M^{-1} \end{bmatrix},$$

$\mathbf{B} = 0$, $\varphi(y) = 0$, then the matrices $\mathbf{A}_o - \mathbf{L}_o \mathbf{C}$, $o \in \{m, M\}$ are cooperative and asymptotically stable (assumption 2 is satisfied). For the parameters

$m_m = .25$, $m_M = .33$; $c_m = 0.08$, $c_M = 0.12$, $c = 0.1$; $\boldsymbol{\theta}_m = [0.5\ 0\ 0\ 0.5]^T$, $\boldsymbol{\theta}_M = [2\ 1\ 1\ 2]^T$, $\boldsymbol{\theta} = [1\ 0.5\ 0.5\ 1.3]^T$,

$$m(t) = 0.5(m_M^{-1} - m_m^{-1})(1 + 0.1(t - 0.5t_k)/[1 + 0.1|t - 0.5t_k|]) + m_m^{-1} + 0.05\sin(3t), \quad M(t) = m_M^{-1} + m_m^{-1} - m(t),$$

where $t_k = 100$ is the simulation time interval, the results of the parameter $\boldsymbol{\theta}$ interval estimation are shown in Fig. 6. The estimates provided by the state observer (22) are plotted in Fig 7. □

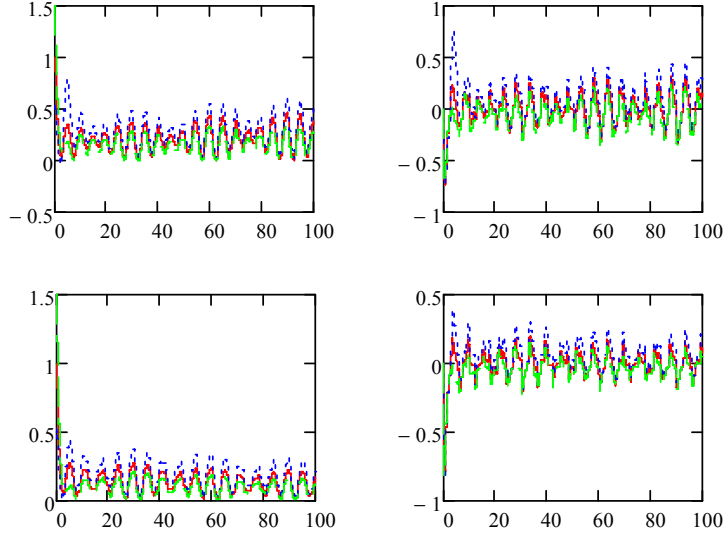

Fig. 7. Upper and lower bounds for the state vector in (24).

**R e m a r k  6**. The requirement imposed in theorems 1−3 on initial conditions $\xi_o(0)$, $\zeta_o(0)$, $\boldsymbol{\Omega}_o(0)$, $\hat{\boldsymbol{\theta}}_o(0)$, $o \in \{m, M\}$ are not restrictive and can be skipped, that may result in additional transients in the intervals evaluation (for linear stable systems the asymptotic behavior is defined by properties of external inputs). □

**R e m a r k  7**. An advantage of the designed solution is that exponential complexity usual for set-membership parameter estimation is avoided. In [15], [16], [25], the problem is formulated as a Constraint Satisfaction Problem (CSP) involving an ordinary differential equation. The CSP is solved in a rigorous way using branch and bound algorithms. The main particularity of these techniques is that the parameter domain is systematically partitioned at each iteration that makes the complexity exponential with respect to the dimension of the parameter vector. It has been proven that the number of iterations is given by:

$$N = (W([\Theta])/\varepsilon + 1)^q,$$

where $W([\Theta])$ is the width of the domain of the parameter vector $\boldsymbol{\theta}$ (a measure of the set $\Theta$); $\varepsilon$ is a tolerance fixed by the user in order to have a result in a finite time, and $q$ is the dimension of the parameter vector. In addition, it is important to note that each iteration should be solved for all the instants of time $t_j$, where $j \geq 0$ lies in the range of the interval of simulation. This process is known to be time-consuming. This limitation is avoided in our work and the dimension of the proposed observer is $2(2n + n \times q + q)$, that is similar to the Kalman filter. This achievement makes reasonable application of the proposed observer to higher dimensional uncertain nonlinear systems. □

Let us consider application of the proposed set adaptive observers in the fault detection problem.

### 6. Fault detection

The main idea of model-based fault detection and diagnosis is to check whether the behavior of the plant is consis-tent with its fault-free model. Many model-based approaches use estimation of some relevant internal or observed variables to produce fault-indicating signals (residuals), see [7] and [8] for a recent survey.

In this section we assume that in the system (3) the faults appearance is modeled by the vector $\boldsymbol{\theta}$ (the absence of faults corresponds to the case $\boldsymbol{\theta} = 0$). The problem is to detect a significant change of the vector $\boldsymbol{\theta}$ value within minimum amount of time.

*A. Fault detection procedure*

To solve this problem, in [26] it is proposed to use the following set observers:

$$\dot{\boldsymbol{\zeta}}_o = \mathbf{A}_o \boldsymbol{\zeta}_o + \mathbf{B}_o \mathbf{u} + \phi(\mathbf{y}_v) + \mathbf{L}_o(\mathbf{y}_v - \mathbf{C}\boldsymbol{\zeta}_o), \ o \in \{m, M\},$$

that coincide with (12). The observers (12) estimate the interval of the state vector values for the nominal case $\boldsymbol{\theta} = 0$. Under some mild assumptions in this case we have $\hat{\mathbf{y}}_m(t) \leq \mathbf{y}(t) \leq \hat{\mathbf{y}}_M(t)$ for all $t \geq 0$, $\hat{\mathbf{y}}_m = \mathbf{C}\boldsymbol{\zeta}_m$, $\hat{\mathbf{y}}_M = \mathbf{C}\boldsymbol{\zeta}_M$, and a failure of this conditions indicates a fault appearance [4], [26]. The fault detection signal is defined as follows

$$S(t) = s_1(t) \vee ... \vee s_p(t), \ s_i(t) = \begin{cases} 0 & \text{if } \hat{y}_{m,i}(t) \leq y_i(t) \leq \hat{y}_{M,i}(t), \\ 1 & \text{otherwise}, \end{cases} \ i = \overline{1, p}, \tag{25}$$

then $S(t) = 0$ in the nominal case and $S(t) = 1$ if a fault is detected (the symbol $\vee$ is stated for the "logic or"). A method of the smallest detectable fault estimation for the observers (12) is also discussed in [4], [26].

What new can be added to this procedure with application of (12)−(14) and (22)? Firstly, let us stress that (12) are incorporated in the adaptive set observers, therefore the indicator (25) can be still verified. Secondly, the observers (12)−(14) provide the interval estimation for the fault vector $\boldsymbol{\theta}$ directly, that allows us to generate the additional fault indicator signal as follows:

$$D(t) = d_1(t) \vee ... \vee d_q(t), \ d_j(t) = \begin{cases} 0 & \text{if } \hat{\theta}_{m,j}(t) \leq 0 \leq \hat{\theta}_{M,i}(t), \\ 1 & \text{otherwise}, \end{cases} \ j = \overline{1, q}. \tag{26}$$

Under conditions of theorems 1 and 2 (exchanging indexes $m$, $M$ probably) a separation of the signal (26) from zero indicates a fault appearance, while the variables $\hat{\boldsymbol{\theta}}_m$, $\hat{\boldsymbol{\theta}}_M$ evaluate the admissible interval of the fault $\boldsymbol{\theta}$ (that can help with the fault isolation). And finally, the observers (22) estimate the state $\mathbf{x}$ values taking into account the interval $[\hat{\boldsymbol{\theta}}_m, \hat{\boldsymbol{\theta}}_M]$, i.e. the condition $\boldsymbol{\xi}_m(t) \leq \mathbf{x}(t) \leq \boldsymbol{\xi}_M(t)$ approves the interval $[\hat{\boldsymbol{\theta}}_m, \hat{\boldsymbol{\theta}}_M]$ and a failure of these bounds implies that either conditions of theorems 1 and 2 are not satisfied or the level of measurement noise/disturbances is very high. Then the third indicating signal can be defined as follows

$$Z(t) = z_1(t) \vee ... \vee z_p(t), \ z_i(t) = \begin{cases} 0 & \text{if } \hat{\psi}_{m,i}(t) \leq y_i(t) \leq \hat{\psi}_{M,i}(t), \\ 1 & \text{otherwise}, \end{cases} \ i = \overline{1, p}, \ \hat{\boldsymbol{\psi}}_m = \mathbf{C}\boldsymbol{\xi}_m, \ \hat{\boldsymbol{\psi}}_M = \mathbf{C}\boldsymbol{\xi}_M. \tag{27}$$

Again, the case $Z(t) = 0$ corresponds to the situation $\hat{\boldsymbol{\theta}}_m \leq \boldsymbol{\theta} \leq \hat{\boldsymbol{\theta}}_M$ and $\boldsymbol{\xi}_m(t) \leq \mathbf{x}(t) \leq \boldsymbol{\xi}_M(t)$, while $Z(t) = 1$ indicates the opposite status. Therefore, the proposed approach consists in a simultaneous verification of the test signals (25)−(27), which gives more tools for fault detection and isolation than the conventional approach based on the set state observers.

Let us demonstrate workability of this approach through a simple application.

*B. State monitoring for three-tanks-system*

As in works [17], [26], [32], [36] consider the three-tank-system presented in Fig. 8 and described by the following equations:

$$S_c \dot{x}_1 = -a_{13}\rho(x_1 - x_3) + u_1 + \theta_1, \quad \rho(x) = sign(x)\sqrt{|x|};$$
$$S_c \dot{x}_2 = -a_{32}\rho(x_3 - x_2) - a_{20}\rho(x_2) + u_2 + \theta_2; \quad (28)$$
$$S_c \dot{x}_3 = a_{13}\rho(x_1 - x_3) - a_{32}\rho(x_3 - x_2) + \theta_3,$$

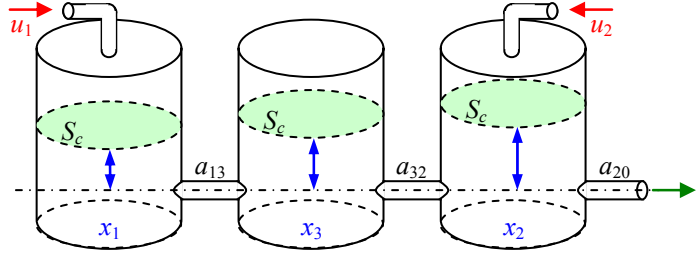

Fig. 8. The structure scheme of the three-tank-system

where the variables $x_i > 0$, $i = \overline{1,3}$ denote the liquids levels in the corresponding tanks, $\mathbf{x} = [x_1 ... x_3]^T$; $u_j$, $j = 1, 2$ are pump flows attached to the tanks 1 and 2, $\mathbf{u} = [u_1 \ u_2]^T$; $S_c$ is the cross section area of the tanks; the tanks are connected via the pipes with outflow coefficients $a_{13} = a_{32}$ and $a_{20}$ is the nominal outflow coefficient, $\mathbf{a} = [a_{13} \ a_{32} \ a_{20}]^T$. The possible actuator faults in the tanks 1 and 2 are modeled by $\theta_1$ and $\theta_2$, the faulty outflow in the tank 3 is described by $\theta_3$, $\boldsymbol{\theta} = [\theta_1 ... \theta_3]^T$.

It is required to design a fault detection system for the model (28). Here we consider two scenarios. In the first one as in [26] we assume that only the variables $x_1$ and $x_2$ are available for measurements and the nominal values of the model (28) parameters ($a_{13}$, $a_{32}$, $a_{20}$ and $S_c$) are given. In this case we do not take into account possible faults in the tank 3 ($\theta_3$ is set to zero). In the second scenario as in [26], [32] all state variables $x_i$, $i = \overline{1,3}$ are accessible for direct measurements, but the model (28) parameters $\mathbf{a}$ belong to some interval of uncertainty, i.e. the real values $\tilde{\mathbf{a}}$ of the model parameters belong to the interval $[r_m\mathbf{a}, r_M\mathbf{a}]$, where the coefficients $r_m$, $r_M$ define admissible deviations of $\tilde{\mathbf{a}}$ from the nominal values $\mathbf{a}$. The parameter $S_c$ is typically known and is not changing during normal operation. In both cases the domain of the state $\mathbf{x}$ values is given, i.e. $\mathbf{x}_m \leq \mathbf{x}(t) \leq \mathbf{x}_M$ for all $t \geq 0$ in the current operating mode.

To apply the approach proposed here we need to transform the system (28) to the form of (3), for this purpose note that $\rho(x)/x = \lambda(x) = |x|^{-0.5}$, then the model (28) can be rewritten as follows:

$$\dot{\mathbf{x}} = \mathbf{A}(\mathbf{x},\mathbf{a})\mathbf{x} + \mathbf{B}\mathbf{u} + S_c^{-1}\boldsymbol{\theta}, \quad (29)$$

$$\mathbf{A}(\mathbf{x},\mathbf{a}) = S_c^{-1}\begin{bmatrix} -a_{13}\lambda(x_1 - x_3) & 0 & a_{13}\lambda(x_1 - x_3) \\ 0 & -a_{32}\lambda(x_3 - x_2) - a_{20}\lambda(x_2) & a_{32}\lambda(x_3 - x_2) \\ a_{13}\lambda(x_1 - x_3) & a_{32}\lambda(x_3 - x_2) & -a_{32}\lambda(x_3 - x_2) - a_{13}\lambda(x_1 - x_3) \end{bmatrix}, \mathbf{B} = S_c^{-1}\begin{bmatrix} 1 & 0 \\ 0 & 1 \\ 0 & 0 \end{bmatrix},$$

that is similar to (3). For the first scenario from (29) we get

$$\mathbf{A}_m(\mathbf{y}) = S_c^{-1}\begin{bmatrix} -a_{13}\lambda(y_1 - x_{M,3}) & 0 & a_{13}\lambda(y_1 - x_{m,3}) \\ 0 & -a_{32}\lambda(x_{m,3} - y_2) - a_{20}\lambda(y_2) & a_{32}\lambda(x_{M,3} - y_2) \\ a_{13}\lambda(y_1 - x_{m,3}) & a_{32}\lambda(x_{M,3} - y_2) & -a_{32}\lambda(x_{m,3} - y_2) - a_{13}\lambda(y_1 - x_{M,3}) \end{bmatrix},$$

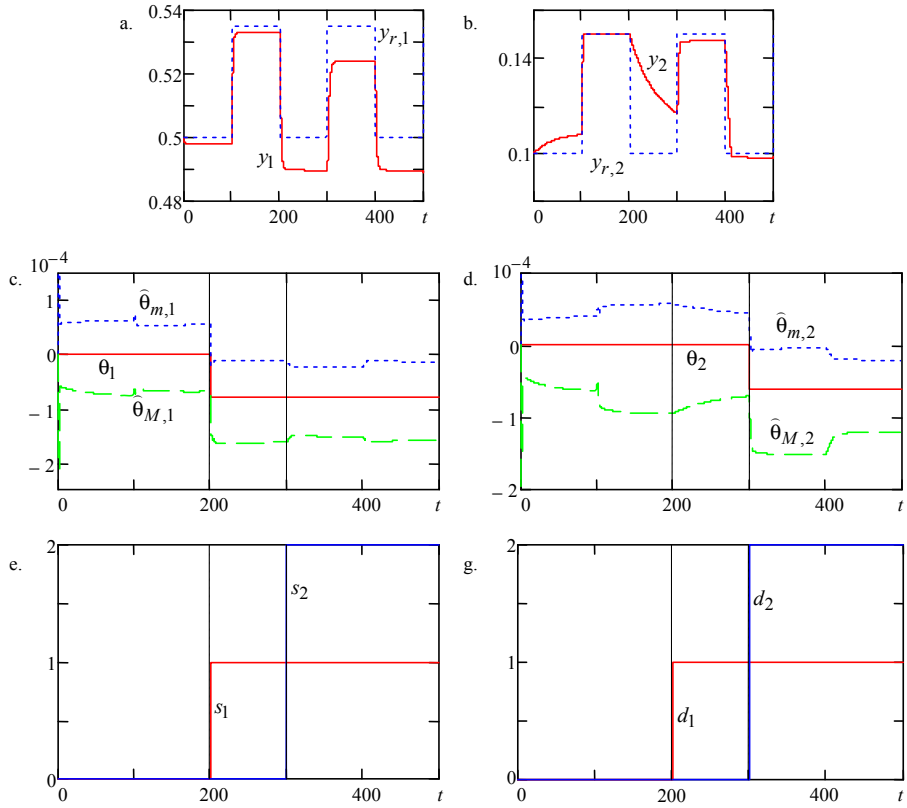

Fig. 9. The results of simulation for the first scenario (without noise): the output $\mathbf{y}$ and its reference $\mathbf{y}_d$ ((a), (b)); $\hat{\boldsymbol{\theta}}_o$ for $o \in \{m, M\}$ ((c), (d)); the fault indicating signals $\mathbf{s}$ and $\mathbf{d}$ ((e), (g)).

$$\mathbf{A}_M(\mathbf{y}) = S_c^{-1} \begin{bmatrix} -a_{13}\lambda(y_1 - x_{m,3}) & 0 & a_{13}\lambda(y_1 - x_{M,3}) \\ 0 & -a_{32}\lambda(x_{M,3} - y_2) - a_{20}\lambda(y_2) & a_{32}\lambda(x_{m,3} - y_2) \\ a_{13}\lambda(y_1 - x_{M,3}) & a_{32}\lambda(x_{m,3} - y_2) & -a_{32}\lambda(x_{M,3} - y_2) - a_{13}\lambda(y_1 - x_{m,3}) \end{bmatrix},$$

$$\mathbf{C} = \begin{bmatrix} 1 & 0 & 0 \\ 0 & 1 & 0 \end{bmatrix}, \quad \mathbf{G} = \mathbf{B}, \quad \mathbf{L}_m = \mathbf{L}_M = \ell \begin{bmatrix} 1 & 0 \\ 0 & 1 \\ 0 & 0 \end{bmatrix}, \quad \ell > 0,$$

and for the second one

$$\mathbf{A}_m(\mathbf{y}) = S_c^{-1} \begin{bmatrix} -r_M a_{13}\lambda(y_1 - y_3) & 0 & r_m a_{13}\lambda(y_1 - y_3) \\ 0 & -r_M[a_{32}\lambda(y_3 - y_2) + a_{20}\lambda(y_2)] & r_m a_{32}\lambda(y_3 - y_2) \\ r_m a_{13}\lambda(y_1 - y_3) & r_m a_{32}\lambda(y_3 - y_2) & -r_M[a_{32}\lambda(y_3 - y_2) - a_{13}\lambda(y_1 - y_3)] \end{bmatrix},$$

$$\mathbf{A}_M(\mathbf{y}) = S_c^{-1} \begin{bmatrix} -r_m a_{13}\lambda(y_1 - y_3) & 0 & r_M a_{13}\lambda(y_1 - y_3) \\ 0 & -r_m[a_{32}\lambda(y_3 - y_2) + a_{20}\lambda(y_2)] & r_M a_{32}\lambda(y_3 - y_2) \\ r_M a_{13}\lambda(y_1 - y_3) & r_M a_{32}\lambda(y_3 - y_2) & -r_m[a_{32}\lambda(y_3 - y_2) - a_{13}\lambda(y_1 - y_3)] \end{bmatrix},$$

$$\mathbf{C} = \mathbf{I}, \quad \mathbf{G} = S_c^{-1}\mathbf{I}, \quad \mathbf{L}_m = \mathbf{L}_M = \ell\mathbf{I}, \quad \ell > 0.$$

Clearly, in both cases the matrices $\mathbf{A}_m$ and $\mathbf{A}_M$ are cooperative and for the chosen gains $\mathbf{L}_m$, $\mathbf{L}_M$ the conditions of assumption 2 are satisfied. Theorem 1 can be applied here due to the matrix $\mathbf{C}$ structure in both scenarios.

For both scenarios the control algorithms are chosen as follows

$$u_1(t, y_1) = \upsilon(-k\rho(y_1 - y_{r,1}(t))), \quad u_2(t, y_2) = \upsilon(-k\rho(y_2 - y_{r,2}(t)) + a_{20}\rho(y_2)), \quad \upsilon(u) = \begin{cases} u & \text{if } u > 0; \\ 0 & \text{otherwise}, \end{cases}$$

where $\mathbf{y}_r(t) = [y_{r,1}(t) \; y_{r,2}(t)]^T$ is the reference signal to be tracked by the state components $x_1$ and $x_2$; $k > 0$ the control gain.

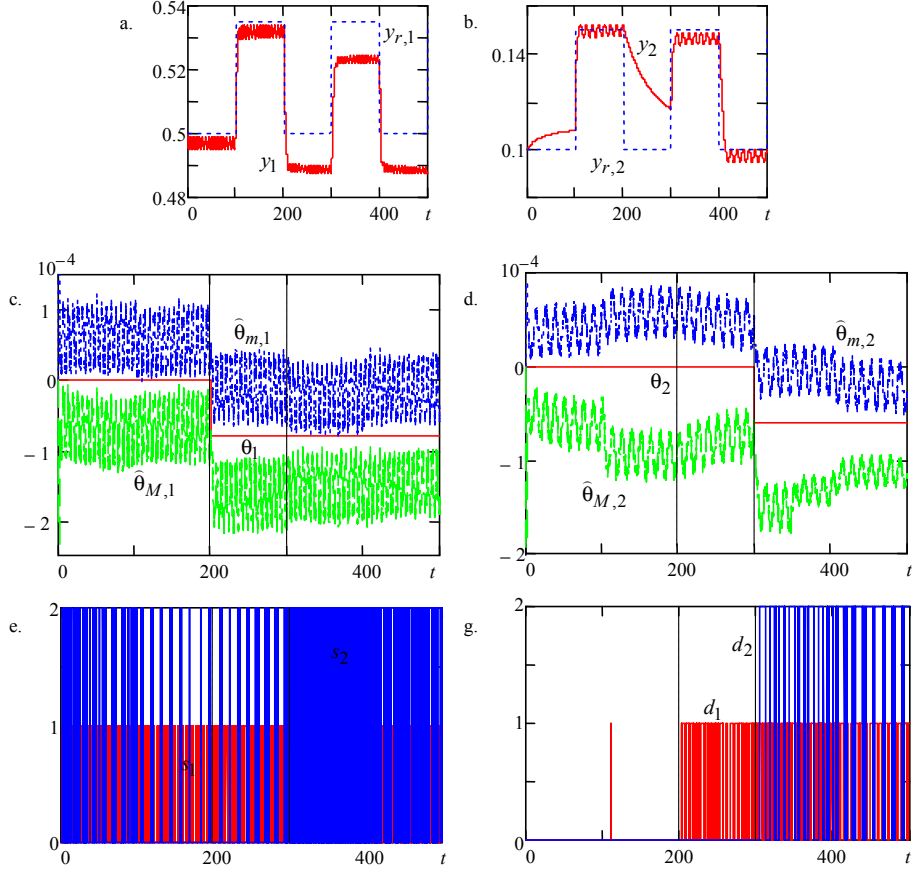

Fig. 10. The results of simulation for the first scenario (with noise): the output $\mathbf{y}$ and its reference $\mathbf{y}_d$ ((a), (b)); $\hat{\boldsymbol{\theta}}_o$ for $o \in \{m, M\}$ ((c), (d)); the fault indicating signals $\mathbf{s}$ and $\mathbf{d}$ ((e), (g)).

The following values of parameters are used for simulation:

$a_{13} = a_{32} = 1.329 \times 10^{-4}$, $a_{20} = 1.772 \times 10^{-4}$, $S_c = 0.0154$, $k = 1.329 \times 10^{-3}$, $\ell = 3$, $\mathbf{x}_m = [0.44\ 0.04\ 0.24]^T$,

$\mathbf{x}_M = [0.56\ 0.16\ 0.36]^T$, $T = 200$, $\mathbf{y}_r(t) = [0.5(1+0.07\mu(t))\ 0.1(1+0.5\mu(t))]^T$, $\mu(t) = \begin{cases} 0 & \text{if } t \bmod T \leq T/2; \\ 1 & \text{otherwise}. \end{cases}$

The initial conditions for the system (28) are chosen as $\mathbf{x}(0) = 0.5(\mathbf{x}_m + \mathbf{x}_M)$.

For the first scenario during the simulation, it is assumed that there are no faults for the first 200 sec, next the fault $\theta_1 = 8 \times 10^{-5}$ appears at the time instant $t_1 = 200$ sec, the fault $\theta_2 = 6 \times 10^{-5}$ appears at $t_2 = 300$ sec (that is 25% and 20% from the maximal control amplitude). The corresponding trajectories are shown in Fig. 9 for the case without noise (the output curves are plotted in Fig. 9,a and b, the graphics of $\boldsymbol{\theta}$, $\hat{\boldsymbol{\theta}}_m$, $\hat{\boldsymbol{\theta}}_M$ are presented in Fig. 9,c and d, the scaled indicating signals $s_i$, $d_i$, $i = 1,2$ are shown in Fig. 9,e and g, the signals $z_i$ are not presented since they are zero during all time of the simulation). The fault detection delays are 0.35 sec and 0.45 sec respectively based on the signals $s_1$ and $s_2$ only. The same trajectories for the case of a stochastic noise presence $|\mathbf{v}(t)| \leq 4.5 \times 10^{-3}$ are plotted in Fig. 10. As it can be seen from the figures 9 and 10, the faults indicating signals (26) are less sensitive to the measurement noise. In this example, based on $d_i$, $i = 1,2$ it is possible to detect faults even in the case of rather noisy measurements.

For the second scenario it is assumed that $r_m = 0.75$, $r_M = 1.25$ and the first two faults appear at similar time in-stants. Additionally, the third fault $\theta_3 = 9 \times 10^{-5}$ appears at $t_3 = 300$ sec. The indicating signals are plotted in Fig. 11 (Fig. 11,a and b represent the case without noise, and Fig. 11, c and d for noisy measurements). Again we note better robustness properties of the signals $d_i$, $i = \overline{1,3}$ comparing them with $s_i$, $i = \overline{1,3}$. The signals $z_i$, $i = \overline{1,3}$ stay zero confirming validity of the indicators $d_i$, $i = \overline{1,3}$. In this scenario, the fault detection delays are 0.52 sec, 0.55 sec and 7.61 sec respectively.

The simulation results confirm good fault detection ability of the proposed adaptive set observers.

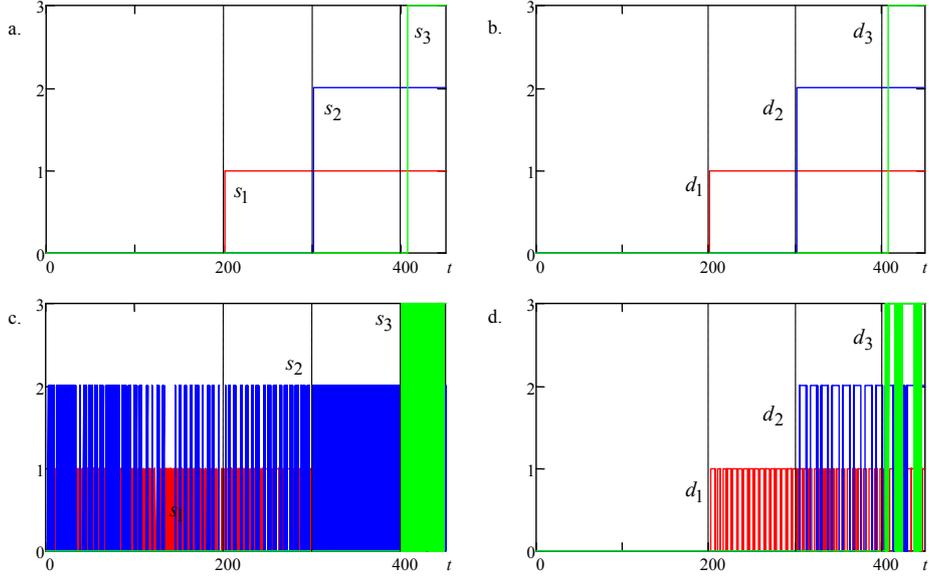

Fig. 11. The results of simulation for the second scenario: the fault indicating signals **s** and **d** without noise ((a), (b)) and with a measurement noise ((c), (d)).

### 7. Conclusion

The basic problem studied by this paper is adaptive observers design for joint parameter and state estimation of nonlinear continuous time systems. Based on a guaranteed LPV approximation, the problem of set observers design for the nonlinear system is reformulated in terms of adaptive observers design problem for LPV ones. The exponential complexity usual for set-membership parameter estimation in nonlinear continuous-time systems is avoided. The complexity of the proposed observer is similar to the Kalman filter and the dimension of the set adaptive observer equations increases proportionally to the parameter $\mathbf{\theta}$ and to the state $\mathbf{x}$ dimensions (the full adaptive set observer dimension is $2(2n + n \times q + q)$). This setting makes possible application of observers for higher dimension uncertain systems.

It is shown that under standard cooperativity assumption imposed on the observer equations, the adaptation loop may be cooperative or competitive depending on additional circumstances. Both competitive and cooperative cases are analyzed and applicability conditions for the adaptive observers are proposed. Moreover, the proposed applicability conditions of the adaptive set observers (presented in Assumption 2) are less restrictive than those corresponding to the conventional adaptive observers (formulated in Assumption 1). Thus, the adaptive set observers can be applied in the cases when the solution of the parameter dependent Lyapunov equation from Assumption 1 is not feasible.

The results of the developed techniques suggest that in the presence of small uncertainties (small deviations of the parameters and the state from their nominal/majorant values) the introduction of adaptive technology may not provide significant improvement in the state estimation. However, if the set of admissible values for the model parameters is largely deviated or under noisy conditions, then the adaptive set observers proposed here could be superior to the al-ready existing solutions.

Finally, it was shown how set adaptive observers can be used to solve the problem of parametric fault detection.